\documentclass[aip, amsmath]{revtex4-1}

\usepackage{graphicx}
\usepackage{dcolumn}
\usepackage{bm}

\usepackage[utf8]{inputenc}
\usepackage[T1]{fontenc}
\usepackage{mathptmx}
\usepackage{etoolbox}
\usepackage{color}
\makeatletter
\def\@email#1#2{%
 \endgroup
 \patchcmd{\titleblock@produce}
  {\frontmatter@RRAPformat}
  {\frontmatter@RRAPformat{\produce@RRAP{*#1\href{mailto:#2}{#2}}}\frontmatter@RRAPformat}
  {}{}
}%
\makeatother
\begin{document}

\preprint{AIP/123-QED}

\title[Bychkov et al., Mangnon-Phonon-Photon Interactions]{Mangnon-Phonon-Photon Interactions in Cubic Ferromagnets in the Vicinity of Orientational Phase Transition: Retrospective and Phenomenological Study}

\author{I.V. Bychkov}
\email{bychkov@csu.ru}
\affiliation{ 
Chelyabinsk State University, Chelyabinsk 454081, Russia
}%

\author{D.A. Kuzmin}%
\affiliation{ 
Chelyabinsk State University, Chelyabinsk 454081, Russia
}%

\author{V.A. Tolkachev}%
\affiliation{ 
Chelyabinsk State University, Chelyabinsk 454081, Russia
}%

\author{V.D. Buchelnikov}
\affiliation{
Chelyabinsk State University, Chelyabinsk 454081, Russia
}%

\author{\framebox{V.G. Shavrov}}
\affiliation{%
Kotelnikov Institute of Radioengineering and Electronics of RAS, Moscow 125009, Russia
}%

\date{\today}

\begin{abstract}
{
This work presents a comprehensive theoretical investigation of magnon-phonon-photon interactions in cubic ferromagnets near orientational phase transitions (OPT). The study focuses on the interplay of magnetoelastic (ME), electromagnetic-spin (EMS), and acoustic-electromagnetic (AEM) interactions in ferromagnetic dielectrics and metals. By deriving dispersion equations for coupled waves, the research reveals how the dynamic properties of magnetically ordered crystals evolve in response to these interactions under varying external magnetic fields and near OPT points.
The ME interaction prominently influences spin and elastic wave dynamics, giving rise to coupled modes such as quasi-magnon and quasi-acoustic waves. Near the OPT, these interactions become dominant, leading to phenomena like the magnetoelastic gap and softening of vibrational modes.
The EMS interaction significantly alters the activation energy and dispersion of quasi-spin and quasi-electromagnetic waves. In ferromagnetic metals, helicons (weakly damped electromagnetic waves) exhibit strong coupling with spin and elastic waves, particularly in the OPT region.
The study identifies conditions for triple resonance among spin, elastic, and electromagnetic waves. Additionally, it explores the frequency-dependent rotation of polarization planes in electromagnetic and magnetoelastic waves, which sharpens near OPT points.
These results provide a deeper understanding of the coupled dynamics in ferromagnetic materials, paving the way for new technological applications in spintronics, signal processing, and advanced magneto-optical devices. The theoretical framework developed here emphasizes the critical role of ME, EMS, and AEM interactions in tailoring wave properties for specific applications, particularly in designing next-generation magnetic and electronic systems. 
}
\end{abstract}

\maketitle

\section{\label{sec:level1}INTRODUCTION}
The nowadays study of magnetoelastic and electromagnetic wave interactions in ferromagnets near orientational phase transitions (OPT) has seen significant advancements, driven by the potential applications in spintronics, sensor technology, and next-generation magnetic devices. Magnetically ordered materials exhibit complex coupling between their spin, elastic, and electromagnetic subsystems, giving rise to phenomena such as coupled spin-elastic waves, the magnetoelastic gap, and topologically protected spin excitations.
Recent research has expanded the theoretical framework of these interactions. The discovery of topological magnon insulators highlights the potential for robust spin wave transport in ferromagnets, which could revolutionize magnon-based computing and data storage \cite{mook2021interaction}. Additionally, the interplay of magnetoelastic effects in thin films and nanostructures has been leveraged to develop devices capable of precisely manipulating wave dynamics for advanced signal processing applications \cite{mazzamurro2020giant}.
Studies have also focused on the Berry curvature of magnetoelastic waves, elucidating their role in non-reciprocal propagation and edge states in magnetic films. These findings pave the way for the design of new materials with anisotropic wave properties, tailored for specific technological uses \cite{okamoto2020berry}.

Despite current interest, the investigations of the dynamical interaction between magnetic, elastic and electromagnetic subsystems in magnetic materials have a long history \cite{vonsovskiĭ1974magnetism,turov2001symmetry,zvezdin2004phase,turov2005new}. The listed interactions play an important role in the formation of many physical properties of magnetically ordered crystals. Each interaction can be characterized by a dimensionless parameter. Usually, except for special cases, the parameters of interaction of subsystems are small. However, when stability in the magnetic subsystem is lost, i.e. in the vicinity of the phase transition, this parameter tends to unity and the role of interaction in the physical properties of magnets increases greatly. In dynamics, each subsystem of the crystal is characterized by its own vibrational modes or elementary excitations (magnons, phonons and photons, respectively). The interaction of subsystems also leads to interaction between elementary excitations and the emergence of coupled vibrations or elementary excitations of a new type. The study builds upon existing theories of magnetoelastic and electromagnetic wave interactions in magnetically ordered crystals. By extending this framework to analyze coupled wave phenomena near OPT, we aim to highlight previously unexplored dynamic effects in ferromagnetic dielectrics and metals.

It is known that in all magnets there are phase transitions of two types - "order - disorder" (Curie and Neel points) and "order - order", the so-called spin-reorientation or orientational phase transitions \cite{vonsovskiĭ1974magnetism,nagaev1988magnets,patashinskii1979fluctuation}. Phase transitions in magnets can be observed both with a change in temperature (spontaneous transitions) and external influences - elastic stresses, electric and magnetic fields (induced transitions). OPT in magnetic crystals are accompanied by a change in the direction of the equilibrium vector of magnetic ordering relative to the crystallographic axes. Approaching the OPT point of a magnet leads to a significant change in the oscillation spectrum and, consequently, the dynamic properties of the crystal. Thus, when a ferro- (FM) or antiferromagnet (AFM) approaches the OPT point, without taking into account the magnetoelastic (ME) interaction, the ferromagnetic mode of oscillations becomes soft; when taking into account the ME interaction, the associated ME waves are soft \cite{dikshtein1974effect, dikshtein1974effects,turov1983broken,bar1984striction,buchel1996magnetoacoustics,patashinskii1979fluctuation,buchel1999relative}.

The first studies on the interaction of the spin and elastic subsystems of magnets, in which the existence of coupled ME waves in FM and AFM was predicted, were carried out in the works of Turov, Irkhin\cite{turov1956vibration}, Akhiezer, Baryakhtar, Peletminsky \cite{akhiezer1958coupled}, Kittel \cite{kittel1958interaction}, Peletminsky \cite{peletminskii1960coupled}. These works were fundamental in the emergence of a new field of physics of magnetically ordered substances – magnetoacoustics. Magnetoacoustics currently finds important applications in electronics and microwave technology \cite{belov1987magnetostriction,goldin1991spin,tacker1975hypersound}.

As noted above, the interactions of subsystems, in particular the ME interaction, are relatively weak interactions. And these interactions manifest themselves most clearly at resonances and in the vicinity of the OPT points. The ME interaction in magnets in the vicinity of the OPT can become predominant or comparable with other interactions due to the decrease in magnetic anisotropy to zero. In this case, the dimensionless parameter of the ME interaction increases sharply and the ME interaction can have a significant effect on the statistical, dynamic, kinetic and other properties of magnets. The study of the influence of magnetic interaction on the properties of magnets began in the 1960s with the works of Tasaki \cite{tasaki1963magnetic}, Mitsumishi \cite{mizushima1966effective}, Iida \cite{iida1964magnetoelastic}. These works gave rise to a new direction in the physics of magnetic phenomena – the study of the effects of strong manifestation of comparatively weak interactions.

In work \cite{tasaki1963magnetic}, during the study of antiferromagnetic resonance in hematite ($ \alpha -Fe_{2}O_{3}$), an effect was experimentally discovered, which in the literature was called the effect of the “magnetoelastic gap” or the “frozen lattice” \cite{cooper1968spin,cooper1968magnetic,cooper1972phenomenological}. From the experiments on antiferromagnetic resonance, it was determined that the resonant frequency is described by the relation

\begin{align} \label{eq:1}
 \omega^2_{0} = \omega^2_{A}+\omega^2_{me}, 
\end{align}

where $\omega_{A}$ is the AFM resonance frequency determined by the external magnetic field and magnetic anisotropy renormalized by magnetostriction, and $\omega_{me}$ is the detected effect of the "magnetoelastic gap" appearing due to the dynamic interaction of the spin subsystem with the crystal lattice. At the point , the OPT $\omega_{A}$ is zeroed and in the dynamics, the ME interaction leads to stretching or compression of the sample in the direction of the antiferromagnetism vector and creates an additional effective field for oscillations of magnetic moments. The magnetization already precesses around this field with a frequency $\omega_{me}$. Dynamic deformations seem to create effective magnetic anisotropy again and the magnetization precesses around this direction - the phase transition seems to disappear in the magnetic subsystem. This is how the elastic subsystem affects the magnetic subsystem near the point and at the OPT point itself.

In \cite{turov1983broken} it was shown that the effect of the "magnetoelastic gap" caused by the ME interaction is common to all magnetically ordered crystals. Theoretical estimates of the ME gap $\omega_{A}$ for rare-earth FMs with easy-plane anisotropy (Dy , Tb) showed that $\omega_{me}/g \sim 10^5$Oe, i.e., has an anomalously large value ( g is the gyromagnetic ratio). These theoretical estimates were confirmed experimentally in experiments on inelastic neutron scattering on spin waves \cite{nielsen1970magnetic,nielsen1970magnon}.
As a result of studies of the ME interaction in crystals with FM and AFM order, it was found that the ME gap represents one side of the phenomenon: it corresponds to only one point of the spectrum (k = 0), the quasi-magnon branch of the spectrum of coupled ME waves. The gap is due to the influence of the elastic subsystem on the magnetic one. The other side of the phenomenon is that the magnetization oscillations also affect the oscillations of the elastic subsystem. Both of these branches of coupled oscillations in the vicinity of the OPT (quasi-magnon and quasi-acoustic) change their appearance compared to the non-interacting branches of oscillations due to the strong interaction of the subsystems. In this case, the interaction of the subsystems turns out to be the strongest, and the magnon gap is completely reduced to the ME gap. The strong distortion of the quasi-acoustic branch is such that as the wave number tends to zero (k$\to$0), the dispersion law for this branch changes from linear to quadratic, this is explained by the fact that magnons “weigh down” phonons. Experimentally, this effect manifests itself as a decrease in the speed of sound with a decrease $\omega_{A}$ in Eq.~(\ref{eq:1}), which can be achieved by changing the magnetic field, temperature or pressure \cite{ozhogin1972easy,seavey1972acoustic,maksimenkov1973antiferromagnetic,tantz1978spin,gorodetsky1970sound,grishmanovskij1974piezomagnetic,gorodetsky1976magnetoelastic}. The softening of the quasiphonon branch also manifests itself in an increase in $\omega_{A}$ sound attenuation near the OPT point \cite{dikshtein1974effect,dikshtein1974effects,turov1983broken,shapira1967absorption,shapira1968ultrasonic,shapira1969ultrasonic,chepurnykh1975some,chepurnykh1975magnetostriction,dikshtein1976spin,buchel1983relaxation,buchel1984magnon}. Far from the OPT, both of these effects (the ME gap and the softening of the phonon mode) are strongly suppressed by the large anisotropy included in the first term in Eq.~(\ref{eq:1}). The ME gap in the spectrum of quasimagnons in the OPT region is associated with antiphase oscillations of the magnetization and the lattice. The analogue of such oscillations are optical oscillations of the lattice. In-phase oscillations of spins and the lattice correspond to quasi-elastic oscillations. The analogue of the latter are acoustic oscillations of the lattice.

Studies of the influence of magnetic interaction on various physical properties were carried out for magnets with different magnetic ordering. The works \cite{dikshtein1974effect,dikshtein1974effects,tasaki1963magnetic,iida1964magnetoelastic,mizushima1966effective,ozhogin1972easy,seavey1972acoustic,maksimenkov1973antiferromagnetic,ozhogin1977effective} were carried out on hematite ($ \alpha -Fe_{2}O_{3}$) , in $FeBO_{3}$ \cite{seavey1972acoustic,tantz1978spin,ozhogin1977effective}, in cubic FM and AFM \cite{berezin1977antiferromagnetic,sokolov1977antiferromagnetic,buchel1983magnetoacoustic}, in orthoferrites \cite{buchel1996magnetoacoustics,buchel1999relative,shapira1967absorption,shapira1968ultrasonic,shapira1969ultrasonic,dan1989dynamic,vitebskij1990anomalous,buchelnikov1994influence,dan1996role,buchel2001observation,dan1998contribution,dan1993energy}, in easy-axis AF near the overturning field ($MnF_{2}, CoF_{2}, Cr_{2}O_{3}$, etc.) \cite{dikshtein1974effects,shapira1967absorption,shapira1968ultrasonic,shapira1969ultrasonic,chepurnykh1975some,chepurnykh1975magnetostriction,dikshtein1976spin,melcher1970magnon}, in easy-axis FM and in ferrites \cite{lee1979ferromagnetic}.

Along with elastic and spin excitations in magnetically ordered crystals (dielectrics, semiconductors and metals) electromagnetic (EM) excitations or EM field oscillations can also exist. In metals and semiconductors, weakly damped spiral EM waves (helicons) can propagate in strong magnetic fields \cite{konstantinov1960possibility,abrikosov2017fundamentals}. The interaction of spin and EM waves in a magnetic dielectric was considered in the work of Akhiezer, Baryakhtar and Peletminskii \cite{akhiezer1958coupled}. Coupled spin and spiral waves propagating along an external magnetic field in a FM metal were considered by Stern and Callen \cite{stern1963helicons} considered the relationship between spin and helical waves in the case of an arbitrary direction of wave propagation. Spector and Casselman \cite{spector1965interaction} considered the interaction of spin waves with Alfvén waves in a FM metal, Baryakhtar, Savchenko and Stepanov \cite{bar1966interaction} investigated the spectrum of coupled plasma, EM and spin waves in FM semiconductors and metals with easy-axis and easy-plane magnetic anisotropy. The interaction of the spin subsystem with the EM field, the so-called electromagnetic-spin (EMS) interaction, leads to a change in the activation value of the quasi-spin branch, i.e. a term of EM nature appears in it – the magnetostatic frequency $\omega_{M}=4\pi g M_{s}$, $M_{s}$ – the saturation magnetization. EMS interaction also leads to a decrease in the phase velocity of EM waves.

The interactions of the spin and elastic, spin and EM subsystems are most clearly manifested in the vicinity of the OPT point, which leads to the emergence of coupled ME and EM waves and a change in the dynamic properties of magnets \cite{buche88fmm}. In particular, these interactions change the activation of quasi-spin oscillations, and at the OPT point it is determined by both ME and EMS interactions

\begin{align} \label{eq:2}
 \omega_{0} = \omega_{me}+\omega_{M}, 
\end{align}

Helicons can also interact with elastic and spin waves \cite{skobov1964theory,blank1968ferromagnetic}. The interaction of helicons with elastic and spin waves far from the OPT in a uniaxial FM metal was studied in \cite{blank1968ferromagnetic}. It was shown there that under certain conditions a triple resonance can be observed in FM, in which all three types of waves are excited.

A comprehensive study of the influence of ME and EMS interactions on the dynamic properties of FM and AFM in the field of OPT stimulated the formulation of new experiments and new theoretical problems related to the interaction of subsystems of magnetically ordered crystals. Experimental and theoretical studies have shown that for a more accurate description of physical properties and phenomena, along with ME and EMS interactions, it is necessary to take into account other interactions existing in magnets and the type of magnetic ordering.

As noted above, the ME and EMS interactions have a significant effect on the propagation of spin, elastic and EM waves in magnets. In the vicinity of the OPT point, these interactions lead to a change in the oscillation spectrum and the occurrence of coupled ME and EM waves. In particular, strong ME coupling in the OPT region should affect the propagation velocity of EM waves, the angle of rotation of the plane of polarization of EM and ME waves \cite{turov1989ground,kaibichev1993rotation,kaibichev1993stability}, as well as the reflection, transmission and absorption of EM waves. However, in the above-mentioned works \cite{akhiezer1958coupled,stern1963helicons,spector1965interaction,bar1966interaction,skobov1964theory,blank1968ferromagnetic}, a complete analysis of the features of the dispersion laws of coupled ME and EM waves in the OPT region was not carried out. This applies to both FM dielectrics and FM metals. The influence of ME, EMS and acousto-electromagnetic interaction (caused by the action of the Lorentz force on the ions of the magnet) on the spectrum of coupled waves in magnets has not been studied previously. Moreover, near the OPT points, the spectrum of coupled ME waves and helicons has not yet been studied.

The aim of the work is a theoretical study of the influence of the interaction of subsystems (magnetic ordered, paramagnetic, elastic, and electromagnetic) on the dynamic properties of magnetically ordered crystals. We present theoretical study of the influence of ME, EMS and AEM interactions on the spectrum of coupled waves in a FM crystal of cubic symmetry located in an external magnetic field coinciding in direction with the magnetization vector \cite{buche88fmm, buche2000fmm}. The case of propagation of coupled waves in FM along the crystallographic axis [001] is considered. Dispersion equations are obtained for an FM dielectric and an FM metal in weak and strong magnetic fields. The features of the behavior of coupled oscillations both far from and near the OPT are investigated. The conditions for fulfilling various types of resonances are determined. The tensor of magnetic susceptibility of FM is obtained taking into account the above interactions. The rotation of the polarization plane of EM and ME waves in an FM dielectric in the OPT region is studied.

The manuscript is organized as follows: Chapter II introduces the energy framework and ground-state properties of ferromagnets. Chapter III develops the dispersion equations for coupled waves, which serve as the foundation for subsequent analyses. Chapter IV applies this framework to ferromagnetic dielectrics, exploring specific conditions under which quadratic dispersion arises. Chapter V extends the analysis to ferromagnetic metals, considering both weak and strong magnetic field regimes. The results are summarized and contextualized in the concluding remarks.

\section{\label{sec:level2}ENERGY AND THE GROUND STATE OF A FERROMAGNET}
For a cubic ferromagnetic crystal, the free energy density $F$ can be written as:

\begin{eqnarray} \label{eq:3}
F = F_{M} + F_{ME} + F_{E} - \textbf{H} \textbf{M} - \frac{1}{2} \textbf{H}_{d}\textbf{M} +\frac{1}{2} \Lambda(\textbf{M}^2-M^2_{0}) 
\end{eqnarray}

Here $F_{M}$ is the magnetic energy density

\begin{eqnarray} \label{eq:4}
F_{M} &=& \frac{1}{2}\alpha M^2_{0} \left( \frac{\partial \textbf{m}}{\partial x_{k}} \right)^2 + K_{1} (m^2_{x} m^2_{y} + m^2_{y} m^2_{z} + m^2_{z} m^2_{x}) + \nonumber\\
& & K_{2}m^2_{x}m^2_{y}m^2_{z}, 
\end{eqnarray}

magnetoelastic energy density

\begin{eqnarray} \label{eq:5}
F_{ME} &=& B_{1} (m^2_{x}u_{xx} + m^2_{y}u_{yy} + m^2_{z}u_{zz}) + 2B_{2} (m_{x}m_{y}u_{xy} + \nonumber \\ 
& & m_{x}m_{z}u_{xz} + m_{y}m_{z}u_{yz}),
\end{eqnarray}

elastic energy density

\begin{eqnarray} \label{eq:6}
F_{E} &=& \frac{1}{2} c_{11} (u^2_{xx} + u^2_{yy} + u^2_{zz}) + c_{12} (u_{xx}u_{yy} + u_{yy}u_{zz} + \nonumber \\ 
& & u_{xx}u_{zz}) + 2c_{44} (u^2_{xy} + u^2_{zy} + u^2_{zx}),
\end{eqnarray}

$\alpha, K, B, c $ are the constants of non-uniform exchange, anisotropy, elasticity;
$\textbf{m}= \textbf{M}/M_{0}$ is a saturation magnetization, $\textbf{M}$ is the magnetization vector, $u_{ij}$ is the strain tensor. 
The fourth and fifth terms in Eq.~(\ref{eq:3}) describe the interaction energy of magnetization with an external magnetic field and a uniform demagnetizing field, respectively (for an infinite sample $\textbf{H}_{d} = 0$). 
The last term in Eq.~(\ref{eq:3}) is introduced to satisfy the condition $ \textbf{M}^2 = \textbf{M}^2_{0}, \Lambda $ is the Lagrange multiplier.

Let us consider an infinite cubic ferromagnet located in an external constant magnetic field $\textbf{H} \parallel \textbf{M}_{0} \parallel \textbf{z}$, the axis $\textbf{z}$ is directed along the crystallographic axis [001]. The solution of the Euler equations for each of the subsystems of the ferromagnet gives its ground state in the form:

\begin{eqnarray} \label{eq:7}
m^0_{x} &=& m^0_{y} = 0, \; m^0_{z} = 1, \; u^0_{xx}=u^0_{yy}=B_{1} \frac{c_{12}}{\Delta_{1}},  \nonumber\\
u^0_{zz} &=& -B_{1} \frac{c_{11}}{\Delta_{1}} -  u^0_{xx}, \; u^0_{ij} = 0, (i \neq j), \\
\Delta_{1} &=& (c_{11}-c_{12})(c_{11} + 2c_{12}), \; \textbf{H}_{0}=\textbf{H}, \nonumber\\
\Lambda &=& H_{0}/M_{0} - 2B_{1}u^0_{zz}/M^2_{0}. \nonumber
\end{eqnarray}

We will study a ferromagnet whose magnetic anisotropy can be described using the first constant of cubic anisotropy $K_{1} < 0$. In this case, with a decrease in the magnitude of the external magnetic field $H$ , at $H = H_{A} =2 |K|/M_{0}$ , a second-order OPT from a collinear state with $\parallel$ to an angular phase with magnetization in a plane of the (110) type $\textbf{z}$ will take place in the ferromagnet $\textbf{M}_{0}$.

\section{\label{sec:level3}SPECTRUM OF COUPLED OSCILLATIONS IN FERROMAGNET}

To obtain the dispersion laws of coupled waves, we will use the system of equations of Landau-Lifshitz, elasticity theory and Maxwell equations
\begin{eqnarray} \label{eq:8}
\frac{\partial \textbf{M}}{\partial t} &=& g \left[\textbf{M} \times \textbf{H}_{eff} \right] + \textbf{R};
\end{eqnarray}
\begin{eqnarray} \label{eq:9}
\rho \frac{\partial^2 u_{i}}{\partial t^2} &=& \frac{\partial \sigma_{ik}}{\partial x_{k}} +\frac{1}{c} \left[\textbf{j} \times \textbf{B} \right]_{i},
\end{eqnarray}
\begin{eqnarray} 
curl \textbf{E} &=& \frac{1}{c} \frac{\partial \textbf{B}}{\partial t}, \; curl\textbf{H} = \frac{4 \pi }{c} \textbf{j} +  \frac{1}{c} \frac{\partial \textbf{D}}{\partial t} \nonumber\\
div \textbf{D} &=& 0, \; div \textbf{B} = 0 \label{eq:10}
\end{eqnarray}

Here $g$ is the gyromagnetic ratio, $\textbf{H}_{eff} = - \frac{\delta F}{\delta \textbf{M}}$ is the effective magnetic field inside the ferromagnet, $R$ is the term taking into account relaxation in the magnetic subsystem, $\textbf{E}, \textbf{H}$ are the electric and magnetic field strengths, respectively, $c$ is the speed of light in a vacuum, $\textbf{j} = \textbf{j}_{disp}+\textbf{j}_{cond}$,  
$\textbf{j}_{disp} = \left( \partial \textbf{D} / \partial t \right) / 4 \pi $ the displacement current density, $\textbf{D} = \widehat{\epsilon} \left( \textbf{E} + \left[ \partial \textbf{u}/ \partial t, \textbf{B}\right] / c \right)$ – the electric field induction, $\widehat{\epsilon}$ is permittivity tensor (for a crystal of cubic symmetry $\widehat{\epsilon} = \epsilon \delta_{ij}$), $\textbf{j}_{cond} = \widehat{\sigma} \left( \textbf{E} + \left[ \partial \textbf{u}/ \partial t, \textbf{B}\right] / c \right)$ is the conductivity current density, $\widehat{\sigma}$ is the electrical conductivity tensor, $\textbf{B} = \textbf{H} + 4 \pi \textbf{M}$ is the magnetic induction, $\rho$ is the FM density of the crystal, $\textbf{u}$ is the displacement vector, $\sigma_{ik} = \frac{1}{2} (\delta_{ik} + 1) \partial F / \partial u_{ik} $ is the stress tensor. In this chapter, we will not take into account relaxation in the magnetic subsystem, i.e., we will put $\textbf{R} = 0$. Let us consider the case of wave propagation along the axis $\textbf{z}, \textbf{k} \parallel \textbf{z},\textbf{k}$   is the wave vector.

In the presence of a magnetic field in a ferromagnetic metal, the off-diagonal (Hall) components should be taken into account in the conductivity tensor $ \widehat{\sigma}$ \cite{abrikosov2017fundamentals}. In the case of strong magnetic fields with a closed Fermi surface, the off-diagonal components of the tensor 
$ \widehat{\sigma}$ exceed the diagonal (dissipative) components

\begin{align} \label{eq:11} 
\widehat{\sigma} = \begin{pmatrix}
  \sigma_{xx} & \sigma_{xy} & 0 \\
  -\sigma_{yx} & \sigma_{yy} & 0 \\
    0 & 0 & \sigma_{zz}
 \end{pmatrix} ,  \nonumber\\
 \sigma_{xy} \gg \sigma_{xx}, \sigma_{yy}; \\ 
 \sigma_{yx} = \frac{(n_{e}-n_{d})ec}{B_{0}} = -\sigma_{xy}. \nonumber
\end{align}

where $n_{e}, n_{d}$ are the densities of electrons and holes in the metal, $e$ is the electron charge. Expression Eq.~(\ref{eq:11}) is valid provided that the spatial variation of the electron distribution function is neglected and provided that the cyclotron frequency $\Omega = eB_{0}/m_{0}$( $m_{0}$ is the effective mass of the electron) exceeds the frequency of electron collisions $\nu$ and the frequency of coupled oscillations $\omega$

\begin{align} \label{eq:12}
kr_{L} \ll 1, \Omega \gg \nu, \omega
\end{align}

where is $r_{L} \approx \nu_{F} / \Omega$ the cyclotron radius, $\nu_{F}$ is the velocity of electrons at the Fermi surface.

We will seek a solution to equations Eq.~(\ref{eq:8})-(\ref{eq:10}) in the form of plane waves propagating along the axis $\textbf{z}$  (proportional to $exp(-i\omega t + ikz)$ ). The dispersion equation of the coupled ME and EM waves has the form

\begin{align} 
(\omega \pm \omega_{e \pm})(\omega_{sk} \pm \omega) (\omega^2-\omega^2_{t})+\omega_{me}\omega^2
(\omega \pm \omega_{e \pm})+ \nonumber\\
\omega_{M} \omega (\omega^2-\omega^2_{t}) - \omega^2_{ea}\omega(\omega_{sk} \pm \omega) - \omega^2_{ea}\omega_{me}\omega=0 \label{eq:13}
\end{align}

here $\omega_{e \pm} = c^2k^2/4 \pi \sigma_{\pm}$ is the dispersion law of non-interacting EM waves of right and left circular polarization, respectively, $\sigma_{\pm} = \sigma_{yx} \mp (i \sigma_{xx} + \epsilon \omega / 4 \pi)$ is the effective conductivity of the ferromagnet, $\omega_{sk} = gM_{0} \alpha k^2 + g(H-H_{A})$ is the dispersion law of non-interacting spin waves, $\omega_{t} = s_{t}k$ is the dispersion law of non-interacting transverse elastic waves, $s_{t}=(c_{44}/\rho)^{1/2}$ is the speed of transverse sound, $c_{44}$ is the modulus of elasticity, $\omega_{me} = g M_{0} \zeta_{me}$ is the frequency of the ME gap, $\zeta_{me} = B^2_{2}/M_{0}c_{44}$ is the dimensionless parameter of the ME coupling, $B_{2}$ is the magnetostriction constant, $\omega_{M} = gM_{0} \zeta_{es}$ is the magnetostatic frequency, $\zeta_{es} = 4\pi$ is the dimensionless parameter of the EMS interaction (or dipole interaction, i.e. the interaction of the variable magnetization of the ferromagnet with the magnetic component of the variable EM field), $\omega^2_{ea} = \zeta_{ea}\omega^2_{t}$, $\zeta_{ea} = B^2_{0}/4 \pi \rho s^2_{t}$ is the dimensionless parameter of the acoustic-electromagnetic (AEM) interaction. At the point of the OPT, $H = H_{A}$ the dispersion law of non-interacting spin waves has a quadratic character: $\omega_{sk} = gM_{0} \alpha k^2$.

In Equation (13), the term $\omega_{e\pm}=c^2k^2/4\pi\sigma_{\pm}$ represents the dispersion of non-interacting electromagnetic waves with right ($+$) and left ($-$) circular polarizations, where $\sigma_{\pm}$ reflects the effective conductivity of the ferromagnet. The spin wave dispersion $\omega_{sk} = gM_0\alpha k^2 + g(H - H_A)$ highlights the influence of exchange stiffness ($\alpha$) and the external magnetic field ($H$). The elastic wave term $\omega_t = k s_t$, where $s_t$ is the speed of transverse sound, describes uncoupled elastic modes. Coupling terms like $\omega_{me}=gM_0\zeta_{me}$ (magnetoelastic interaction) and $\omega_M=gM_0\zeta_{es}$ (electromagnetic-spin interaction) demonstrate the interplay between spin, elastic, and electromagnetic subsystems. Finally, $\omega_{ea}$ incorporates acousto-electromagnetic coupling, arising from Lorentz force effects on ions, which plays a critical role in metals near OPT points.

Often, to study other dynamic properties of ferromagnets, for example, when studying the reflection of EM and elastic waves from the surface of a ferromagnet, it is necessary to know the dynamic magnetic permeability tensor. The dynamic magnetic permeability tensor is obtained from the solution of a system of coupled equations Eq.~(\ref{eq:8})-(\ref{eq:10}) taking into account the ME, EMS and AEM interactions.

The dispersion equation  Eq.~(\ref{eq:13}) allows us to obtain the dispersion laws of coupled ME and EM waves both in FM dielectrics and in FM conductors. Let us consider the cases of a dielectric and a conductor separately.

\section{\label{sec:level4}FERROMAGNETIC DIELECTRIC}

Building on the theoretical framework established in Chapters II and III, this chapter applies the derived dispersion relations to ferromagnetic dielectrics. The magnetoelastic coupling in such materials is well-documented; however, we aim to refine existing models by investigating new conditions under which quadratic dispersion arises, particularly in high-permittivity dielectrics ($\epsilon \approx 10^3$).

For a ferromagnetic dielectric, we can set $\sigma_{xx}=\sigma_{yx} = 0$ and then the frequency $\omega_{e \pm}$ will take the form

\begin{align} \label{eq:14}
\omega_{e \pm} = \mp \frac{c^2k^2}{\epsilon \omega}
\end{align}

Substituting this expression into  Eq.~(\ref{eq:13}) we obtain the following dispersion equation of coupled ME and EM waves for a FM dielectric of cubic symmetry

\begin{align} 
(\omega^2 - \omega^2_{e})(\omega_{sk} \pm \omega) (\omega^2-\omega^2_{t}) + \omega_{me}\omega^2(\omega^2 - \omega^2_{e}) + \nonumber \\
\omega_{M} \omega^2 (\omega^2-\omega^2_{t}) - \omega^2_{ea}\omega^2(\omega_{sk} \pm \omega) - 
\omega^2_{ea}\omega_{me}\omega^2=0 \label{eq:15}
\end{align}

Equation (15) governs the coupled magnetoelastic and electromagnetic wave dynamics in ferromagnetic dielectrics. Here, $\omega_{e} = ck/\sqrt{\epsilon}$ is the dispersion of electromagnetic waves, while $\omega_{me}$ and $\omega_{ea}$ highlight the contributions of magnetoelastic and acousto-electromagnetic interactions, respectively. The derived solutions ($\omega_{1-5}$) provide insights into how these couplings alter wave behavior, particularly near OPT points. For instance, $\omega_{1}$ corresponds to quasi-spin waves with significant activation energy due to electromagnetic-spin interactions, while $\omega_{4,5}$ describe quasi-electromagnetic modes that exhibit quadratic or linear dispersion under varying conditions.

At the point of the OPT $H=H_{A}$ and in the long-wave region, $\omega_{e} \ll \omega_{M}$ the dispersion equation  Eq.~(\ref{eq:15}) has solutions

\begin{eqnarray} \label{eq:16} 
\omega_{1} &=& \omega_{M}+ \omega_{me}, \nonumber \\
\omega_{2,3} &=& \left( \frac{\omega_{me} (\omega^2_e + \omega^2_{ea})+ \omega_{M} \omega^2_{t}}
{\omega_{me} + \omega_{M}} \right)^{1/2}, \\
\omega_{4,5} &=& \frac{1}{2} \frac{\omega^2_{e}\omega^2_{t}}{\omega_{me}(\omega^2_{e}+\omega^2_{ea})+\omega_{M}\omega^2_{t}} \times \nonumber \\
& &\left[ \left( 1+\frac{4\omega_{sk} [ \omega_{me} (\omega^2_{e}+\omega^2_{ea}) + \omega{M} \omega^2_{t}] }{\omega^2_{e}\omega^2_{t}} \right)^{1/2}\pm 1\right] \nonumber 
\end{eqnarray}

\begin{figure}[h!]
\includegraphics[width=1\linewidth]{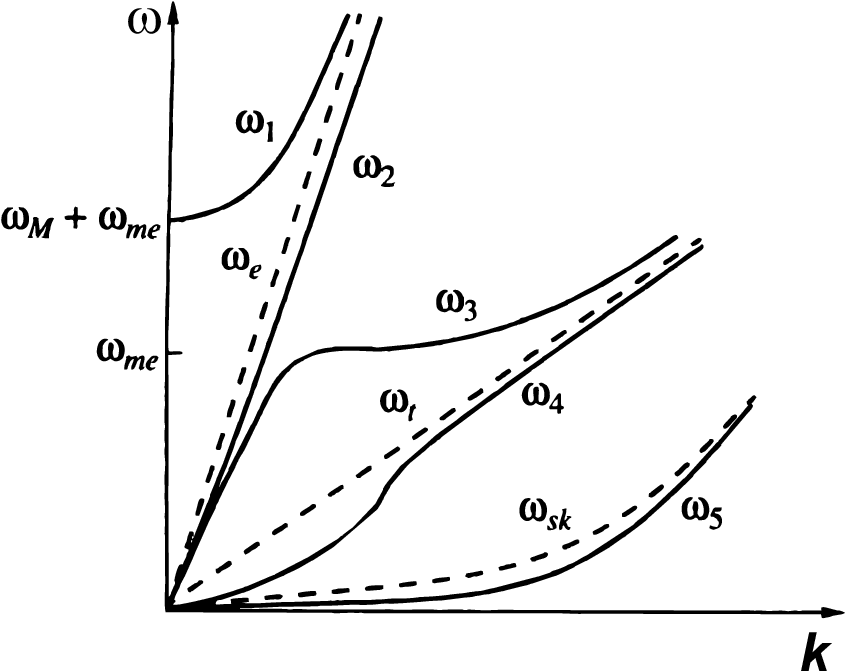}
\caption{\label{fig:fig.1} Schematic spectrum of the boundaries of magnetoelastic and electromagnetic waves in a cubic ferrodielectric at the location of the OFP ($H_{0} = H_{A}$). Here and in  Fig.~\ref{fig:fig.2}, the dotted lines indicate the non-interacting directions of electromagnetic ($\omega_{e}$), elastic ($\omega_{t}$) and spin ($\omega_{sk}$) waves; solid curves are the direction of electromagnetic and magnetoelastic waves ($\omega_{1}-\omega_{5}$).}
\end{figure}

It is evident that the spectrum of coupled oscillations consists of five branches  Fig.~\ref{fig:fig.1}. One of the branches, the quasi-spin ($\omega_{1}$), has an activation determined by the interaction of spin waves with EM waves (term $\omega_{M}$), as well as with elastic waves (term $\omega_{me}$). Branches $\omega_{2}$ and $\omega_{3}$ have a linear dispersion law. In the case of, $\omega_{me}\omega^2_{e} \ll \omega_{M}\omega^2_{t}$ these solutions describe quasielastic waves, and in the opposite case, quasi-electromagnetic waves. The last two branches have a quadratic dispersion law. When $\omega_{me}\omega^2_{e} \ll \omega_{M}\omega^2_{t}$ these branches are quasielectromagnetic, and when the opposite inequality is true, they are quasielastic. Thus, it follows from  Eq.~(\ref{eq:16}) that in ferromagnetic dielectrics, when taking into account the interaction of EM and ME waves in the region of small wave numbers, not only quasielastic waves, but also quasi-electromagnetic waves can have a quadratic dispersion law. However, the condition under which such a situation can occur in FM dielectrics is very strict. For this, it is necessary that the relationship between the characteristics of the magnet be fulfilled

\begin{align} \label{eq:17}
\zeta_{me} c^2/4\pi \epsilon < s^2_{t}
\end{align}

For example, for typical values of the constants $B_{2}=1 \cdot 10^6 erg/cm^3$ and $M_{0} = 2 \cdot 10^3 G$, $s_{t} = 3 \cdot 10^5 cm/s$, condition  Eq.~(\ref{eq:17}) can be satisfied only at $\epsilon \sim 1 \cdot 10^3$. The permittivity can take such large values either in ferromagnetic materials or in the frequency range of its anomalous dispersion. It also follows from  Eq.~(\ref{eq:16}) that the influence of the AEM interaction on the spectrum of coupled waves in FM dielectrics is very insignificant, since the ratio $(\omega_{ea}/ \omega_{e})^2 = B^2_{0}/4\pi \rho c^2 \ll 1$ even in superstrong magnetic fields up to 200 kOe and more.

In the region of wave numbers, $\omega_{t} \ll \omega_{me} \ll \omega_{M} \ll \omega_{e}$ the solutions of the dispersion equation  Eq.~(\ref{eq:15}) can be written in the form

\begin{eqnarray} \label{eq:18}
\omega_{1,2} &=& \omega_{e}(1 \pm \omega_{M}/\omega_{e}), \nonumber \\
\omega_{3} &=& \omega_{me}, \\
\omega_{4,5} &=& \frac{1}{2} \frac{\omega^2_{t}}{\omega_{me}} \left[ \left( 1 + 
\frac{4 \omega_{sk} \omega_{me}}{\omega^2_{t}}\right)^{1/2} \pm 1 \right] \nonumber
\end{eqnarray}

\begin{figure}[h!]
\includegraphics[width=1.0\linewidth]{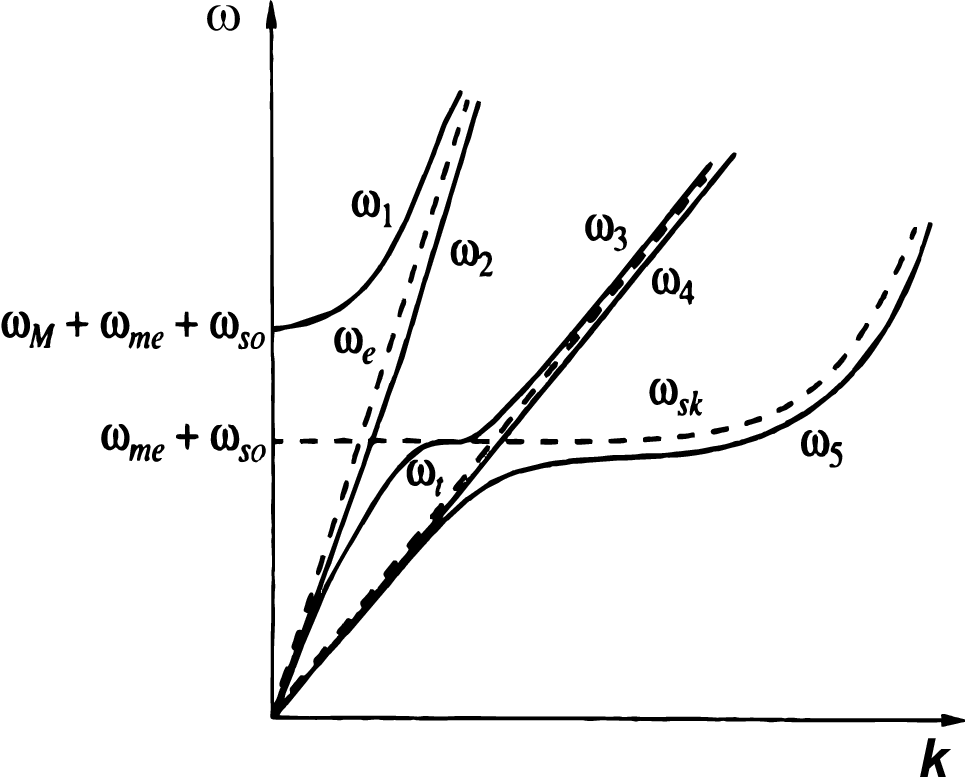}
\caption{\label{fig:fig.2} Schematic spectrum of coupled magnetoelastic and electromagnetic waves in a cubic ferromagnetic dielectric far from the OPT point. The solid lines indicate quasi-spin ($\omega_5$), quasi-acoustic ($\omega_{3,4}$), and quasi-electromagnetic ($\omega_{1, 2}$) wave branches. The frequency $\omega_{S0} = \omega_{Sk}(k = 0) $ represents the characteristic softening frequency of spin wave modes near OPT due to magnetoelastic interactions.}
\end{figure}

It follows that in the region of wave numbers $k \gg \omega_{M} (\epsilon)^{1/2}/c$ (at typical values of constants $g \sim 2 \cdot 10^7 Oe^{-1}s^{-1}$, $M_{0} \approx 10^3 Gs$, $c = 3 \cdot 10^{10} cm/s$ this inequality corresponds to the region $k \ll 10 cm^{-1}$ ) the interaction of ME waves with EM waves can be neglected and ME waves (branches $\omega_{3,4,5}$) can be considered separately from EM waves (branches $\omega_{1,2}$). Thus, in the region of wave numbers $\omega_{M} \ll \omega_{e}$ the dispersion law of quasi-spin waves is determined by the ME coupling, and only quasi-elastic branches can have a quadratic dispersion law. The dispersion law of quasi-electromagnetic waves differs only slightly from the dispersion law of non-interacting EM waves. The dependence of the frequencies of coupled ME and EM waves on the wave number far from the OPT point is schematically shown in  Fig.~\ref{fig:fig.2}, where the dispersion curves $\omega_{1, 2}$ and $\omega_5$ correspond to quasi-electromagnetic and quasi-spin waves, respectively, while $\omega_{3,4}$ represent quasi-acoustic modes. The frequency $\omega_{S0} = \omega_{Sk}(k = 0)$ characterizes the onset of mode softening due to strong magnetoelastic coupling.

By varying the applied fields, the frequency of the magnons can be modified and brought into resonance with the phononic modes in the frequency region where spin and acoustic waves interact (branches $\omega_{3}$ and $\omega_{5}$). Experimentally \cite{novotny2010, rodriguez2016} it was observed that the Fourier amplitude spectrum exhibits avoided crossing. This is in stark contrast to the SAW-FMR behavior. There are two peaks at each field value observed at and around the crossover field, which are the hybridization of the magnon and phonon modes \cite{holanda2018, man2017}. The energy splitting of hybridized modes can be tuned by various external degrees of freedom, which is especially desirable for the development of reconfigurable magnonic devices \cite{vogel2015, krawczyk2014}. Recently, the coupling strength has been significantly improved an unprecedented coupling strength of ~8 in a Galfenol nanograting. \cite{godejohann2020} It was found that tuning the magnon mode to a quasi-transverse phonon mode led to a clear band splitting, suggesting strong hybridization, while tuning the magnon frequency to a quasi-longitudinal phonon mode did not lead to observable hybridization. In addition, a mode strength comparable to that of magnons and phonons is also necessary for strong hybridization. In summary, such a strongly coupled magneto-mechanical system may provide the possibility of developing more efficient transducers between the magnonic and phononic systems \cite{berk2019}. 

The findings in this chapter extend the known results by specifying the conditions for quadratic dispersion in ferromagnetic dielectrics, emphasizing the interplay between magnetoelastic coupling and dielectric properties. These results set the stage for Chapter V, where similar interactions are analyzed in ferromagnetic metals with more complex conductivity characteristics.

\section{\label{sec:level5}FERROMAGNETIC METAL IN WEAK MAGNETIC FIELD}

Expanding upon the analysis of ferromagnetic dielectrics in Chapter IV, this chapter explores coupled wave phenomena in ferromagnetic metals. The introduction of electrical conductivity and the interplay between diagonal and off-diagonal components of the conductivity tensor ($\sigma_{xx}, \sigma_{yx}$) adds a new layer of complexity. The goal is to identify how these interactions influence the dispersion laws, particularly under weak and strong magnetic field regimes.

In this work, 'weak magnetic field' refers to conditions where $H < H_{crit}$, leading to the dominance of diagonal components ($\sigma_{xx} \approx ne ^2 \tau / m$, where $n$ is an electron concentration, $e$ is an electron charge, $m$ is an effective electron's mass, $tau$ is a relaxation time) in the conductivity tensor. 'Strong magnetic field' corresponds to $H \gg H_{crit}$, where off-diagonal components ($\sigma_{xy}, \sigma_{yx} \approx \sigma_{xx} / \omega_c \tau $, where $\omega_c = eB/mc$ is a cyclotron frequency) prevail. For instance, in ferromagnetic metals, $H_{crit} \approx m c / e \tau \approx 100 $ Oe, and the transition between these regimes occurs around $H \approx 10^3 - 10^4$ Oe.

In this case, we can put $\epsilon=0$ and $\sigma_{xx} \gg \sigma_{yx}$. Then the frequency $\omega_{e \pm}$ will be written as

\begin{align} \label{eq:19}
\omega_{e \pm} = \omega_{e} = \mp ic^2 k^2 / 4 \pi \sigma_{xx},
\end{align}
and the dispersion equation  Eq.~(\ref{eq:13}) will take the form

\begin{align} 
(\omega \pm \omega_{e} )(\omega_{sk} \pm \omega)(\omega^2 - \omega^2_{t} ) + \omega_{me} \omega^2 
(\omega \pm \omega_{e}) + \nonumber \\
\omega \omega_{M} (\omega^2 - \omega^2_{t} ) - \omega\omega^2_{ea}(\omega_{sk} \pm \omega) - \omega \omega^2_{ea}\omega_{me} = 0. \label{eq:20}
\end{align}

In the absence of interaction between the subsystems ($\omega_{me} = 0$, $\omega_{ea} = 0$, $\omega_{M} = 0$), equation  Eq.~(\ref{eq:20}) describes the propagation of spin, elastic, and EM waves in a magnetic metal. According to  Eq.~(\ref{eq:19}), EM waves are strongly damped – they have a skin effect. When taking into account the interaction between the subsystems, the presence of a skin effect for EM waves will cause damping of ME waves as well.

At the point of the OPT $H = H_{A}$ and at small wave numbers, the spectrum of coupled oscillations of a ferromagnetic metal can be approximately written as

\begin{eqnarray} \label{eq:21}
\omega_{1} &=& \omega_{M} + \omega_{me} + \omega_{sk} + \frac{\omega_{e}\omega_{M}}
{\omega_{M}+\omega_{me}} + \frac{\omega^2_{t}(\omega_{M} \zeta_{ea} + \omega_{me})}
{(\omega_{M}+\omega_{me})^2};  \nonumber \\
\omega_{2,5} &=& \mp \frac{\omega_{e}\omega_{sk}}{\omega_{M} + \zeta_{ea}\omega_{me}}; \\ 
\omega_{3,4} &=& \omega_{t} \left(1-\frac{\omega_{me} (1- \zeta_{ea})}{\omega_{M} +  \omega_{me}} \right)^{1/2} \mp \nonumber \\
& & \frac{\omega_{e}\omega_{me} (\omega_{M}+\omega_{me}) - \omega^2_{t} (\zeta_{ea} \omega_{M} + \omega_{me})}{2(\omega_{M} + \omega_{me})^2}. \nonumber
\end{eqnarray}

\begin{figure}[h!]
\includegraphics[width=1.0\linewidth]{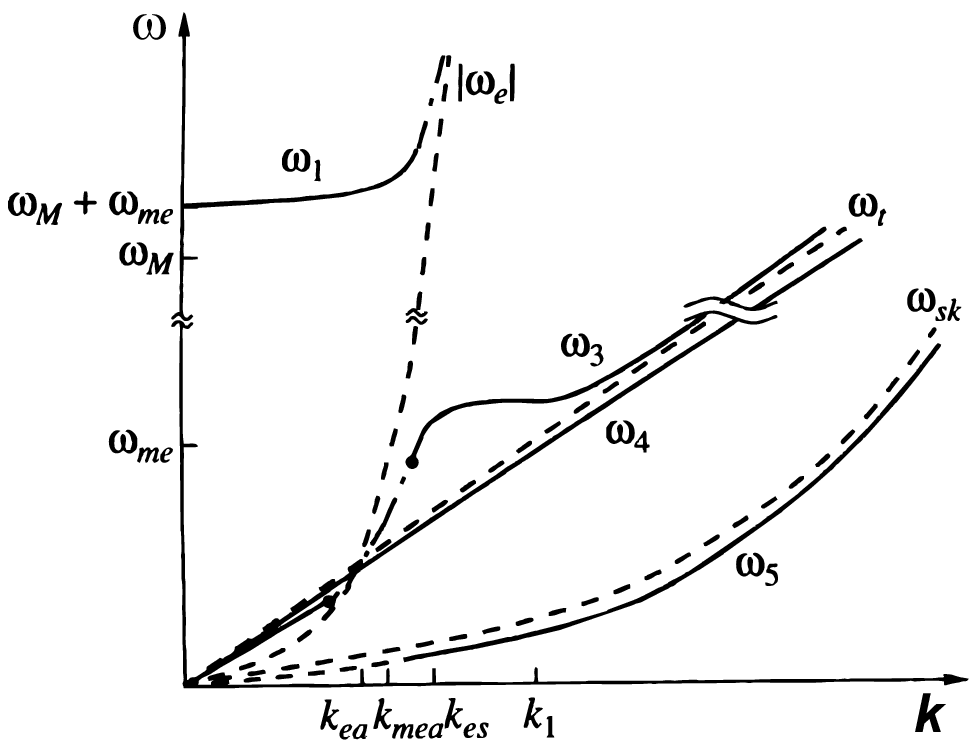}
\caption{\label{fig:fig.3} Schematic spectrum of coupled ME and EM waves in a cubic ferromagnetic metal in weak magnetic fields at the point of the OPT at $k_{ea} < k_{mea}$ . Here and in  Fig.~\ref{fig:fig.4}, the dotted lines indicate non-interacting branches of electromagnetic ($|\omega_{e}|$), elastic ($\omega_{t}$) and spin ($\omega_{sk}$) waves; solid curves are branches of coupled electromagnetic and magnetoelastic waves ($\omega_{1}-\omega_{5}$). Dash-and-dot curves are regions of strong attenuation of branches of coupled electromagnetic and elastic waves. Wavenumbers$k_{ea}, k_{mea}, k_{es}$ and $k_{1}$ correspond to the intersection points of frequencies $|\omega_{e}|$ and $\omega_{t}$, $|\omega_{e}|$ and $\omega_{me}$, $|\omega_{e}|$ and $\omega_{M}$, $\omega_{me}$ and $\omega_{t}$, respectively.}
\end{figure}

\begin{figure}[h!]
\includegraphics[width=1.0\linewidth]{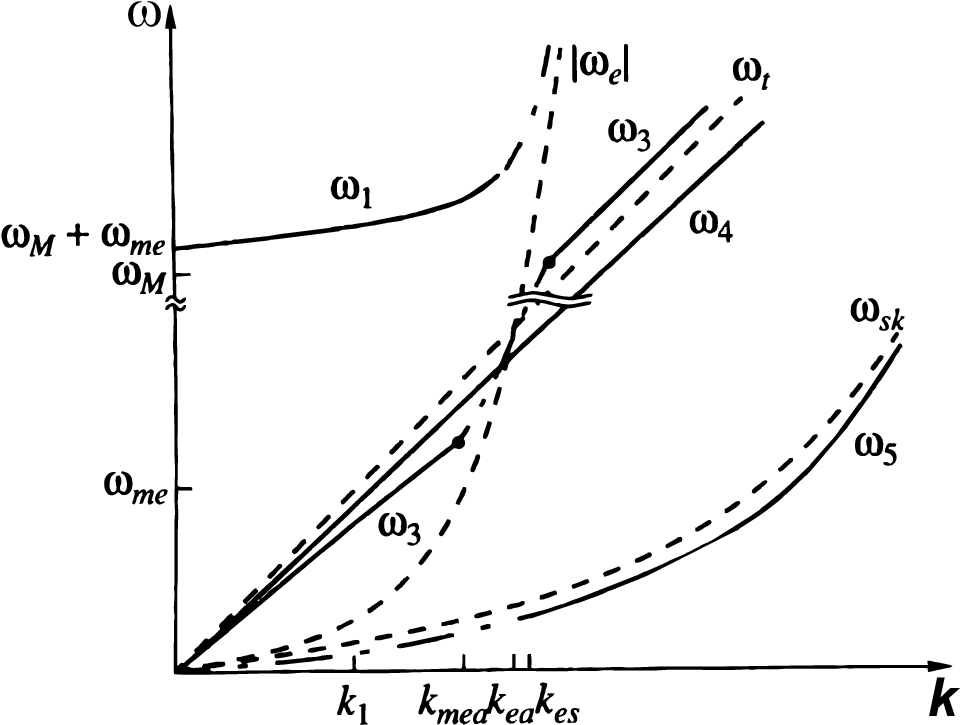}
\caption{\label{fig:fig.4} Schematic spectrum of coupled magnetoelastic and electromagnetic waves in a cubic ferromagnetic metal in weak magnetic fields at the OPT point at $k_{ea}>k_{mea}$.}
\end{figure}

From this it is evident that, indeed, taking into account the interaction of ME waves with non-propagating EM waves leads to the fact that the quasi-elastic and quasi-spin waves become damped (the terms in  Eq.~(\ref{eq:21}) are proportional to $\omega_{e}$  Eq.~(\ref{eq:19})). The dispersion law of the quasi-elastic branches $\omega_{3,4}$ in the metal at $k \to 0$ remains linear with respect to the wave number, and it differs only slightly from the dispersion law of elastic waves due to the smallness of the ratio $\omega_{me}/\omega_{M}$ in the FM  Fig.~\ref{fig:fig.3},  Fig.~\ref{fig:fig.4}. The activation of the quasi-spin branch $\omega_{1}$ , as in the FM dielectric, is determined by the interaction of spin waves with EM and elastic waves. Unlike quasi-elastic waves, the dispersion law of non-propagating quasi-electromagnetic waves $\omega_{2,5}$ changes dramatically. At the point of the OPT, $H=H_{A}$ it depends on the wave number as $k^4$. This leads to a decrease in the thickness of the skin layer in the FM metal by several orders of magnitude. Indeed, from  Eq.~(\ref{eq:21}) we obtain the following expression for the effective thickness of the skin layer in the FM metal at the OPT point $\delta_{eff} = (c^2 \alpha /16 \pi^2 \sigma_{xx} \omega)^{1/4}$.

For example, at $ \alpha \sim 10^{-12} cm^2, \sigma_{xx} \sim 10^{17} s^{-1}$ and $\omega \sim 10^8 s^{-1}$ the effective thickness of the skin layer is $\delta_{eff} \sim 3 \cdot 10^{-4} cm$, while in a nonmagnetic metal it will be equal to $\delta_{eff} \sim 3 \cdot 10^{-3} cm$, i.e. an order of magnitude greater.

Analysis of the dispersion equation  Eq.~(\ref{eq:21}) for other values of wave numbers leads to the following results. For wave numbers $k \sim k_{ea}=4 \pi \sigma_{xx} s_{t}/c^2$ (that is, in the region of intersection of the dispersion laws of evanescent EM and elastic waves $|\omega_{e}| = \omega_{t}$; for typical values of constants $\sigma_{xx} \approx10^{17} s^{-1}, s_{t} \approx 3 \cdot 10^5 cm/s$ the value $k_{ea}$ is $ \approx 10^3 cm^{-1}$), the right-polarized quasielastic waves $\omega_{3}$ become strongly damped (the region corresponding to the dashed-dotted section in  Fig.~\ref{fig:fig.3} and  Fig.~\ref{fig:fig.4}). For them, the skin effect occurs. If the relation is satisfied in the FM metal $k_{ea} < k_{mea} = (4 \pi \sigma_{xx} \omega_{me} / c^2)^{1/2}$ (this is possible, for example, at $B_{2} \approx 5 \cdot 10^7 erg/cm^3, c_{44} \approx 10^{11} erg/cm^3, M_{0} \approx 100 G$, when the wave number is $k_{mea} \approx 2 \cdot 10^3 cm^{-1}$), then at $k > k_{mea}$ branch $\omega_{3}$ becomes quasi-spin, and at $k=k_{1} > \omega_{me}/s_{t}$ it again turns into a right-polarized quasi-elastic branch  Fig.~\ref{fig:fig.3}. In the opposite case, it is always quasi-elastic  Fig.~\ref{fig:fig.4}. In the region of wave numbers $|\omega_{e}|>\omega_{M}$ (or $k>k_{es} = (4 \pi \sigma_{xx}\omega_{M} / c^2)^{1/2}$), the branch becomes non propagating. In this range of wave numbers, the branch $\omega_{1}$ with the dispersion law $\omega_{5} \approx \omega_{sk}$ is propagating $\omega_{5}$  Fig.~\ref{fig:fig.3} and  Fig.~\ref{fig:fig.4}. The dispersion laws of the branches $\omega_{3}, \omega_{4}$ and $\omega_{5}$ describe coupled ME waves. If again in FM metal the relation $k_{ea} < k_{mea} = (4 \pi \sigma_{xx}\omega_{me} / c^2)^{1/2}$  Fig.~\ref{fig:fig.3} is fulfilled, then the branch $\omega_{5}$ at $k_{1}>k>k_{mea}$ describes quasi-elastic waves, and at $k>k_{1}$ it has a quasi-spin character  Fig.~\ref{fig:fig.3}. In the opposite case, its character is always quasi-spin  Fig.~\ref{fig:fig.4}. Thus, in the region of wavenumbers, $k>k_{ea}$ we can neglect the AEM and EMS interactions and consider coupled waves as ME. In this case, for $k_{ea}<k_{mea}$ in the region of wave numbers, $k_{1}>k>k_{mea}$ the dispersion law of right-polarized quasi-elastic waves will depend quadratically on $k$.

This chapter reveals that in ferromagnetic metals, the interaction between magnetoelastic waves and non-propagating electromagnetic waves leads to significant damping effects and novel dispersion behaviors, such as quartic dispersion ($k^4$). These effects are further amplified in strong magnetic fields, where helicons interact strongly with quasi-elastic and quasi-spin waves. These findings complement the results from Chapter IV, demonstrating the broader applicability of the theoretical framework.

\section{\label{sec:level6}FERROMAGNETIC METAL IN STRONG MAGNETIC FIELD}

In a strong magnetic field, the main contribution to the frequency $\omega_{e \pm}$ is made by the off-diagonal component of the conductivity tensor $\sigma_{yx}$. In this case, the frequency $\omega_{e \pm}$ takes the form

\begin{align} \label{eq:22}
\omega_{e \pm} = \omega_{e} = c^2 k^2 / 4 \pi \sigma_{yx},
\end{align}

and the dispersion equation for coupled waves will coincide with equation  Eq.~(\ref{eq:20}). Expression  Eq.~(\ref{eq:22}) describes the propagation of non interacting weakly damped spiral EM waves – helicons – in a FM metal.

\begin{figure}[h!]
\includegraphics[width=1.0\linewidth]{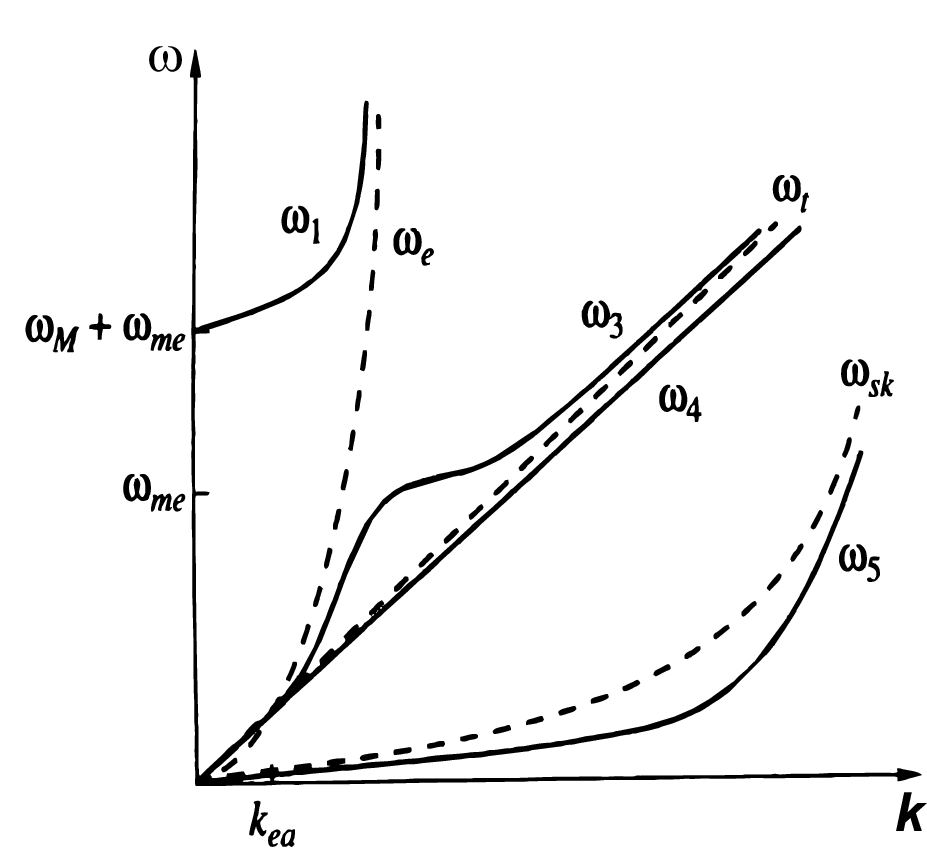}
\caption{\label{fig:fig.5} Schematic spectrum of coupled magnetoelastic and electromagnetic waves in a cubic ferromagnetic metal in strong magnetic fields at the OPT point at $\omega_{M} \ll \omega_{e}(k_{ea})$. Here and in  Fig.~\ref{fig:fig.6}, the dotted line indicates non-interacting branches of electromagnetic ($\omega_{e}$), acoustic ($\omega_{t}$) and spin ($\omega_{sk}$) waves; solid curves are branches of coupled electromagnetic and magnetoelastic waves ($\omega_{1} - \omega_{5}$). Wavenumber $k_{ea}$ corresponds to the frequency intersection point $\omega_{e}$ and $\omega_{t}$.}
\end{figure}

At the OPT point $H =H_{A}$ and at small wave numbers ($k \to 0$), the solutions of the dispersion equation  Eq.~(\ref{eq:20}) are expressed by formulas  Eq.~(\ref{eq:21}). As in the previous case, at $k\to 0$, the dispersion law of quasi-elastic oscillations $\omega_{3,4}$ depends linearly on the wave number  Fig.~\ref{fig:fig.5}. The quasi-spin branch $\omega_{1}$ still has an activation determined by the dipole and ME interactions. Since right-hand-polarized helicons are propagating, the interaction of elastic and spin waves with EM waves does not lead to their attenuation in large magnetic fields. At the OPT point, $H =H_{A}$ the dispersion law of quasi-helicons changes most significantly $\omega_{5}$. From  Eq.~(\ref{eq:21}) it follows that it depends on the wave number as $k^4$.

In addition to the intersection of non-interacting branches at a point, $k=0$ there may also be an intersection of branches at $k \neq 0$ . The helicon and elastic branches intersect at a point $k_{ea} = s_{t}/D_{h}$, where $D_{h} = cB_{0}/(4 \pi Ne)$, $N = n_{e} - n_{d}$. The dispersion laws of the spin and elastic branches intersect at a point $k_{me} = s_{t}/D_{m}$, where $D_{m} = gM_{0} \alpha$. At the point of the OPT at, $k \neq 0$ there is no interaction between the helicons and spin waves. For typical values of the FM constants \cite{turov1983broken, blank1968ferromagnetic} $ M_{0} \sim 10^3 G, N \sim 6 \cdot 10^{22} $cm$^{-3}, g = 2 \cdot 10^7 $s$^{-1}$Oe$^{-1}, \alpha \cdot 10^{-12} $cm$^2 , s_{t} \sim 3 \cdot 10^5 $cm/s we obtain the following values of the wave numbers for the intersection points of non-interacting branches:  $k_{ea} \sim 4 \cdot 10^9 /B_{0} $cm$^{-1}, k_{me} \sim 10^7 $cm$^{-1}$. It follows that at the point of the OPT it makes sense to consider only the phonon-helicon resonance, since the magnetoacoustic resonance occurs at the limit of applicability of the macroscopic approach. The phonon-helicon resonance at $B_{0} \sim 4 \cdot 10^4 Gs$ should be observed at $k_{ea} \sim 4 \cdot 10^5$ cm$^{-1}$, which corresponds to frequencies of $\sim 10^{10}$ s$^{-1}$.

For the acoustic-helicon resonance (i.e. at $\omega=\omega_{e}=\omega_{t}$) the following solutions are obtained from the dispersion equation  Eq.~(\ref{eq:20}). At $\omega_{M} \gg \omega_{e}(k_{ea})$  Fig.~\ref{fig:fig.5}

\begin{eqnarray} \label{eq:23}
\omega_{1,2} &=& \omega_{e}+\omega_{M}+\omega_{me}, \nonumber \\
\omega_{3} &=& \omega_{e} \left[1+ \frac{\zeta_{ea}(\omega_{me}-\omega_{e})}{2\omega_{M}+\omega_{me}} \right], \\
\omega_{4} &=& \omega_{e} \left[1- \frac{\omega_{me}(1-\zeta_{ea}/2)}{\omega_{M}} + \frac{\omega_{e}\zeta_{ea}}{2\omega_{M}} \right]. \nonumber
\end{eqnarray} 

\begin{figure}[h]
\includegraphics[width=1.0\linewidth]{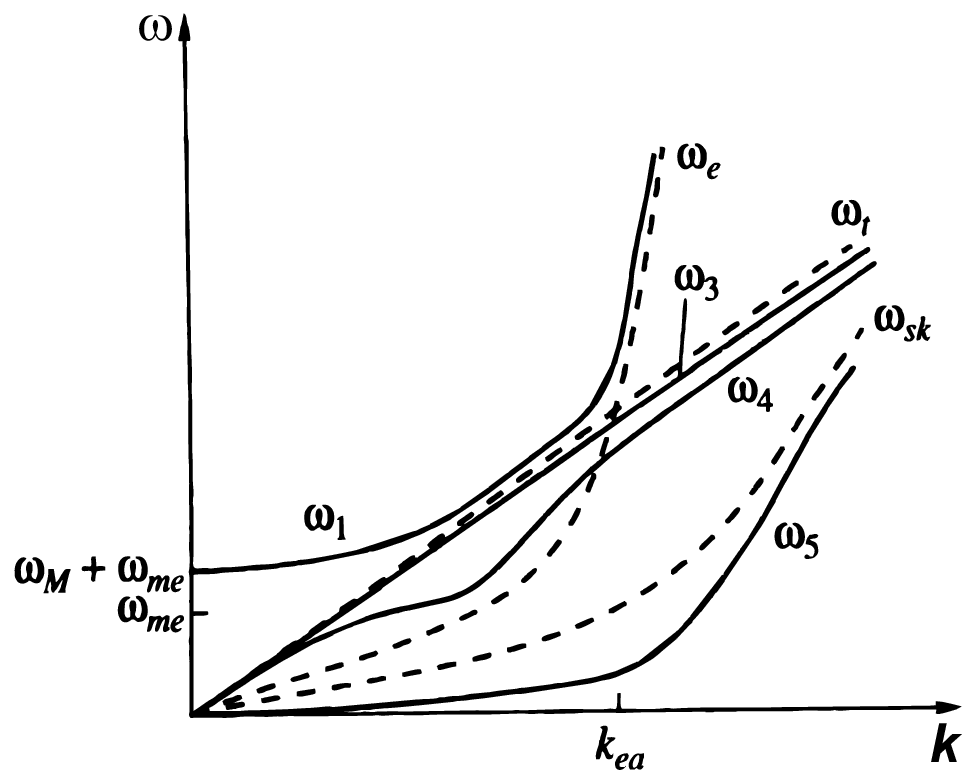}
\caption{\label{fig:fig.6} Schematic spectrum of coupled magnetoelastic and electromagnetic waves in a cubic ferromagnetic metal in strong magnetic fields at the OPT point at $\omega_{M} \gg \omega_{e}(k_{ea})$.}
\end{figure}

In case of the reverse inequality $\omega_{M} \ll \omega_{e}(k_{ea})$  Fig.~\ref{fig:fig.6}

\begin{eqnarray} \label{eq:23a}
\omega_{1,2} &=& \omega_{e} \left[1- \frac{\omega_{me}}{2\omega_{e}} + \frac{\zeta_{ea}}{4}\right], \\
\omega_{3,4} &=& \omega_{e} \left\{1+ \frac{2 \omega_{M}+\omega_{me}}{2\omega_{e}} \pm \frac{1}{2} \left[\frac{(2\omega_{M}+\omega_{me})^2}{4\omega^2_{e}}+4\zeta_{ea} \right]^{\frac{1}{2}} \right\} \nonumber
\end{eqnarray}

The lower branch of coupled oscillations ($\omega_{5}$) is not shown here. From  Eq.~(\ref{eq:23}) it follows that at acoustic-helicon resonance the helicons and elastic waves of the same (right) polarization interact most strongly. The shift of resonance frequencies for them is of the order of  $\omega_{M} \approx 3 \cdot 10^{11}$ s$^{-1}$. Elastic waves of left polarization interact weakly with helicons. Their dispersion law differs slightly from the dispersion law of non interacting waves due to the smallness of the AEM interaction parameter $\zeta_{ea}$.

The nature of the dispersion laws of coupled waves changes as a function of the wave number as follows  Fig.~\ref{fig:fig.5} and  Fig.~\ref{fig:fig.6}. In the case of, $\omega_{e}(k_{ea}<\omega_{M})$ the branch $\omega_{1}$ is first quasi-spin, and then quasi-helicon. The branch $\omega_{3}$ at small wave numbers is quasi-elastic with a linear dispersion law, then its nature changes to quasi-helicon with a quadratic dependence on the wave number.

With a further increase in the wave number, this branch acquires a quasi-spin character with a dispersion law that is practically independent of $k$ and determined by the ME coupling, and then it again acquires a quasi-elastic character. The branch $\omega_{5}$ at small wave numbers is quasi-helicon with a dispersion law $\sim k^4$. Then it first has a quasi-elastic character, and then a quasi-spin character with a change in the dependence of the dispersion law on the wave number from $k^4$ to $k^2$. The branch $\omega_{4}$ at all wave numbers is quasi-elastic. Thus, in this case, at wave numbers determined from the inequality $\omega_{e}>\omega_{me}$, branches $\omega_{3}, \omega_{4}$ and $\omega_{5}$ describe coupled ME waves and the influence of EM waves on them in this region of wave numbers can be neglected. The change in the character of the coupled waves at can $\omega_{e}(k_{ea}) >\omega_{M}$ be easily traced in  Fig.~\ref{fig:fig.6}.

It is of interest to study the behavior of the dispersion laws of coupled waves in strong fields far from the OPT point ($H=H_{A}$). Far from the OPT point, the elastic branch of oscillations has one point of intersection with the helicon branch (at $k \neq 0$) and two points of intersection with the spin branch in the case $s^2_{t} > 4 \omega_{s0} D_{m}$, where $\omega_{s0} = g(H-H_{A})$, and has no intersection when the inverse inequality is satisfied. The second point of intersection of the elastic branch with the spin branch at $s^2_{t} > 4 \omega_{s0} D_{m}$ still lies in the region of the applicability limit of the macroscopic description and can be excluded from consideration here. In the case  of $s^2_{t} > 4 \omega_{s0} D_{m}$, of $D_{h} < D_{m}$, only an acoustic-helicon resonance can take place. For standard values of the characteristics of the FM metal, the latter condition will be satisfied at $B_{0} < $ 100 Gs. At $D_{h}>D_{m}$, i.e. in the region of magnetic fields $B_{0}> $ 100 Gs, three resonances take place: acoustic-helicon, spin-helicon, and magnetoacoustic. The wave numbers corresponding to these resonances are determined by the following formulas:

\begin{eqnarray} 
k_{ea} &=& \frac{s_{t}}{D_{h}}, k_{es} = \left(\frac{\omega_{s0}}{D_{h}-D_{m}}\right)^\frac{1}{2}, \nonumber \\
k_{me} &=& \frac{s_{t}}{2D_{m}} \left[1-\left(1-\frac{4\omega_{s0}D_{m}}{s^2_{t}}\right)^\frac{1}{2}\right] \label{eq:24}
\end{eqnarray}

All resonance wave numbers  Eq.~(\ref{eq:24}) depend on the external magnetic field included in the quantities $D_{h}$ and $\omega_{s0}$. As a result, a triple resonance can occur in a FM metal \cite{blank1968ferromagnetic}, when the frequencies of elastic, spin, and EM waves coincide. The condition for the coincidence of frequencies can be written as:

\begin{align} \label{eq:25}
\omega_{H}D^2_{h}+\omega_{A}D^2_{h}-D_{h}s^2_{t} + D_{m}s^2_{t} = 0
\end{align}

This equation is a third-degree equation with respect to the magnetic field strength. With the first anisotropy constant of a cubic FM $|K| \sim $ 10 $^5 $ erg/cm$^3$ and standard values of other FM parameters given above, for the magnetic field strength at which triple resonance should be observed from  Eq.~(\ref{eq:25}) we obtain the following expression

\begin{align} \label{eq:26}
H_{r} \approx \left[\frac{4 \pi N e s^2_{t}}{gc}\right]^{1/2} \sim  \text{7 kOe}
\end{align}

This field corresponds to a wave number $k \sim$ 10$^5$ cm$^{-1}$ or frequency $\omega \sim 10^{10} s^{-1}$.
At, $H_{0}<H{r}$ the wave number $k_{me}$ corresponding to magnetoacoustic resonance lies to the left of the wave number $k_{es}$ corresponding to spin-helicon resonance. The relationship between the wave numbers $k_{me}$ is also $k_{ea}$ the opposite in this case. At, $H_{0}>H_{r}$ the relationship between the wave numbers  Eq.~(\ref{eq:24}) changes to the inverse of the above conditions.

At $s^2_{t}<4\omega_{s0}D_{m}$ and $D_{h}<D_{m}$ only an acoustic-helicon resonance can take place, and at $D_{h}>D_{m}$ - acoustic-helicon and spin-helicon resonances. In this case, triple resonance in FM metal is impossible.
Note that the influence of fluctuations on the spectrum of coupled oscillations near the OPT was not considered here. It is known that in the case of the OPT, the region near the transition in which fluctuations can significantly affect the propagation of waves is negligibly small and is not observed in experiments.

\section{\label{sec:level7}ROTATION OF THE PLANE OF POLARIZATION SOUND AND ELECTROMAGNETIC WAVES IN A FERROMAGNETIC DIELECTRIC}

The dispersion equation  Eq.~(\ref{eq:15}) (in the case of $\omega_{ea}=0$) has $k$ several roots relative to the wave number $\omega \ll \omega_{s0}$ at a given frequency. In the case of frequencies $\omega$, there will be four such roots. Two of them correspond to left- and right-polarized EM waves

\begin{align} \label{eq:27}
k^{\pm}_{em} = \frac{\omega}{c} \left[\epsilon \frac{\omega_{s0}+\omega_{me}+\omega_{M} \pm \omega}{\omega_{s0}+\omega_{me} \pm \omega}\right]^{1/2} 
\end{align}

the other two are left- and right-polarized quasi-elastic waves

\begin{align} \label{eq:28}
k^{+}_{ke} = \frac{\omega}{s} \left[\frac{\omega_{s0}+\omega_{me}+\omega}{\omega_{s0}+ \omega}\right]^{1/2}, s^2=c_{44}/\rho, \nonumber \\
k^{-}_{ke} = \begin{cases}
   \frac{\omega}{s} \left[\frac{\omega_{s0}+\omega_{me}-\omega}{\omega_{s0}- \omega}\right]^{1/2}, & \quad  \omega_{s0} \gg \omega, \\ 
    \left[\frac{\omega-\omega_{s0}}{g \alpha M_{0}}\right]^{1/2}, & \quad  \omega_{s0} \ll \omega.
  \end{cases}
\end{align}

The obtained formulas indicate that the speed of the left- and right-polarized EM and ME waves in cubic FM differ. Because of this difference in speeds, rotation of the plane of polarization of EM and ME waves occurs. The following formula is valid for the specific angle of rotation of the plane of polarization

\begin{align} \label{eq:29}
\chi = \phi/L= \frac{1}{2}|k^--k^+|
\end{align}

where $L$ is the length of the sample. The magnitude of the angle of rotation of the plane of polarization depends resonantly on the frequency $\omega$. The greatest rotation is observed for EM waves at a frequency of $\omega \sim \omega_{s0}+ \omega_{me}$, and for ME waves – at a frequency  of $\omega \sim \omega_{s0}$. Note that when approaching the OPT, the resonant frequency for both EM and ME waves decreases. At the transition point itself, the resonant frequency for EM waves is equal to the ME gap in the spectrum of spin waves, and for ME waves it is zero. The rotation angle for both types of waves near the OPT increases sharply.

\section{\label{sec:level8}FEATURES OF COUPLED ELECTROMAGNETIC AND MAGNETOELASTIC WAVES IN LIMITED MEDIA}

In limited samples at  $\omega_{ea}=0$ the dispersion equation of coupled ME and EM waves in a FM dielectric  Eq.~(\ref{eq:15}) will remain practically unchanged. It is mainly the frequencies of non-interacting waves that change. Now the wave number $k$ will take a discrete series of values and, consequently, the frequencies $\omega_{e}, \omega_{sk}, \omega_{t}$ will also take discrete values. The allowed values $k$ are determined by the dimensions of the sample. For example, in the case where the sample has the shape of a sphere, the wave number for the lower waves is approximately expressed by the formula

\begin{align} \label{eq:30}
k=2\pi/R.
\end{align}

where $R$ is the radius of the sphere. EM waves will interact with ME waves only if $\omega_{ea} \leq \omega_{sk}$. This inequality, together with condition  Eq.~(\ref{eq:30}), allows us to estimate the dimensions of the sample at which EM waves do not affect the spectrum of ME oscillations.

\begin{align} \label{eq:31}
R \ll 2\pi c/ \sqrt{\epsilon}\omega_{s0}
\end{align}

For example, at $ \epsilon \sim 10, \omega_{S0}/2\pi \sim 10^{10} s^{-1}, R \leq $ 0.6 cm. The estimate obtained is in good agreement with the results of the experimental work \cite{danshin86mhd}.

\section{\label{sec:level9}CONCLUDING REMARKS}

Foundational studies, such as those by Turov and Shavrov \cite{turov1983broken}, investigated magnetoelastic coupling in weakly interacting systems, while Buchel’nikov et al. \cite{buche88fmm} extended these ideas to complex coupled wave phenomena near OPT points. Unlike these earlier works, which focused on specific interactions, this study introduces a unified framework that incorporates ME, EMS, and AEM interactions. This approach provides a more comprehensive understanding of coupled wave dynamics in ferromagnetic dielectrics and metals.

The developed theoretical framework provides a unified approach to analyzing coupled wave phenomena in ferromagnetic materials. Chapters IV and V demonstrate the applicability of this framework to dielectrics and metals, respectively, revealing distinct dynamic behaviors under different material properties and external field conditions. Together, these findings emphasize the versatility of the proposed approach and its potential to guide future experimental and technological developments.

Key insights include the identification of dispersion features for coupled waves in both ferromagnetic dielectrics and metals. Near the OPT, strong ME interactions lead to phenomena such as the magnetoelastic gap, mode softening, and the emergence of coupled quasi-magnon and quasi-acoustic modes. In ferromagnetic metals, helicons exhibit strong coupling with spin and elastic waves, particularly at specific resonance conditions. The rotation of polarization planes for electromagnetic and magnetoelastic waves, sharply resonant near the OPT, further highlights the intricate coupling mechanisms.

It is shown that in the region of small wave numbers, in FM dielectrics, when taking into account the interaction of ME and EM waves, not only quasi-elastic waves, but also quasi-electromagnetic waves can have a quadratic dispersion law. However, such a situation can be observed only at large values of the permittivity of a magnet, of the order of $10^3$. In another case, quasi-electromagnetic waves have a linear dispersion law. Taking into account the AEM interaction does not have a special effect on the spectrum of coupled waves even in super-strong fields. In the region of large wave numbers, the interaction of ME waves with EM can be neglected, and only quasi-elastic waves can have a quadratic dispersion law in this range.

In metals, at the point of the OPT, in the presence of weak magnetic fields, taking into account the interaction of ME waves with non propagating EM waves leads to the fact that in the spectrum of oscillations of quasi-spin and quasi-elastic waves, sections appear in which these waves become strongly damped, and the dispersion law of quasi-electromagnetic waves changes from quadratic to $k^4$, which leads to a significant decrease in the skin layer in the FM metal.
In large magnetic fields, quasi-spin and quasi-elastic waves interact with propagating helicons of right polarization. Due to this interaction, at the point of the OPT, the dispersion law of quasi-helicon waves becomes proportional to the fourth power of the wave number.

The emergence of quadratic ($k^2$) and quartic ($k^4$) dispersion relations highlights the unique interplay of magnetoelastic and electromagnetic interactions in ferromagnetic materials. Quadratic dispersion is associated with systems where magnetoelastic coupling dominates, enabling robust propagation of quasi-acoustic modes. In contrast, quartic dispersion reflects stronger subsystem coupling, such as spin-lattice interactions, and is particularly prominent in ferromagnetic metals near OPT. These features distinguish the studied systems from conventional ferromagnets and open avenues for applications in precision magnonic devices and advanced signal processing technologies.

The magnitude of the angle of rotation of the plane of polarization in a FM dielectric resonantly depends on the frequency. The greatest rotation is observed for EM waves at a frequency of $\sim \omega_{s0}$, and for ME waves – at frequencies of $\sim \omega_{s0}+\omega_{me}$. When approaching the OPT, the resonant frequency for both EM and ME waves decreases. At the transition point, the resonant frequency for EM waves is equal to the ME gap, and for ME waves is zero. The angle of rotation for both types of waves near the OPT increases sharply.

EM waves will interact with ME waves only in the case when $\omega_{e} \leq \omega_{sk}$. This inequality allows us to estimate the sample sizes at which EM waves do not affect the spectrum of ME oscillations; for typical FM, the linear size of the sample is several millimeters.

These findings deepen the understanding of how ME, EMS, and AEM interactions influence wave propagation and resonances in ferromagnets. They provide a robust theoretical framework for exploring new experimental designs and advancing applications in spintronics, magneto-optical devices, and signal processing technologies. Future studies could extend this work to include fluctuation effects near the OPT and explore coupling phenomena in more complex magnetic systems.

Although this work is primarily theoretical, its predictions can be validated experimentally. For instance, the rotation of the polarization plane near OPT points could be tested using the magneto-optical Kerr effect, where polarization rotation depends sharply on frequency near the magnetoelastic gap. Similarly, triple resonance conditions in ferromagnetic metals, as predicted here, could be explored using inelastic neutron scattering or Brillouin light scattering, which are well-suited to probing coupled spin, elastic, and electromagnetic waves in magnetically ordered materials.

In conclusion, the theoretical groundwork established here underscores the critical role of weak interactions in shaping the dynamic behavior of ferromagnetic systems, offering valuable insights for both fundamental research and technological innovation. Our findings are particularly relevant for the development of next-generation spintronic devices, where control over magnon and phonon propagation is critical. For example, the quadratic dispersion in ferromagnetic dielectrics can be utilized in magnonic waveguides for efficient energy transfer, while quartic dispersion in metals is promising for high-frequency magneto-optical devices, such as filters or resonators operating in the gigahertz range. The control of wave dispersion through magnetoelastic interactions can enable more efficient magnonic waveguides, critical for data transmission in spintronic circuits. Moreover, the predicted polarization rotation effects near OPT points could be applied in advanced magneto-optical sensors, which require high sensitivity in the gigahertz frequency range. These results align with ongoing efforts to develop energy-efficient signal processing devices and pave the way for next-generation hybrid magnon-photon systems.

\section*{Acknowledgments}

Research was curried out with support of Ministry of Science and Higher Education of the Russian Federation (075-00187-24-00).

\section*{Data Availability Statement}

Data sharing is not applicable to this article as no new data were created or analyzed in this study.

\nocite{*}
\bibliography{aipsamp}

\providecommand{\noopsort}[1]{}\providecommand{\singleletter}[1]{#1}%
\begin{thebibliography}{87}%
\makeatletter
\providecommand \@ifxundefined [1]{%
 \@ifx{#1\undefined}
}%
\providecommand \@ifnum [1]{%
 \ifnum #1\expandafter \@firstoftwo
 \else \expandafter \@secondoftwo
 \fi
}%
\providecommand \@ifx [1]{%
 \ifx #1\expandafter \@firstoftwo
 \else \expandafter \@secondoftwo
 \fi
}%
\providecommand \natexlab [1]{#1}%
\providecommand \enquote  [1]{``#1''}%
\providecommand \bibnamefont  [1]{#1}%
\providecommand \bibfnamefont [1]{#1}%
\providecommand \citenamefont [1]{#1}%
\providecommand \href@noop [0]{\@secondoftwo}%
\providecommand \href [0]{\begingroup \@sanitize@url \@href}%
\providecommand \@href[1]{\@@startlink{#1}\@@href}%
\providecommand \@@href[1]{\endgroup#1\@@endlink}%
\providecommand \@sanitize@url [0]{\catcode `\\12\catcode `\$12\catcode `\&12\catcode `\#12\catcode `\^12\catcode `\_12\catcode `\%12\relax}%
\providecommand \@@startlink[1]{}%
\providecommand \@@endlink[0]{}%
\providecommand \url  [0]{\begingroup\@sanitize@url \@url }%
\providecommand \@url [1]{\endgroup\@href {#1}{\urlprefix }}%
\providecommand \urlprefix  [0]{URL }%
\providecommand \Eprint [0]{\href }%
\providecommand \doibase [0]{http://dx.doi.org/}%
\providecommand \selectlanguage [0]{\@gobble}%
\providecommand \bibinfo  [0]{\@secondoftwo}%
\providecommand \bibfield  [0]{\@secondoftwo}%
\providecommand \translation [1]{[#1]}%
\providecommand \BibitemOpen [0]{}%
\providecommand \bibitemStop [0]{}%
\providecommand \bibitemNoStop [0]{.\EOS\space}%
\providecommand \EOS [0]{\spacefactor3000\relax}%
\providecommand \BibitemShut  [1]{\csname bibitem#1\endcsname}%
\let\auto@bib@innerbib\@empty
\bibitem [{\citenamefont {Mook}\ \emph {et~al.}(2021)\citenamefont {Mook}, \citenamefont {Plekhanov}, \citenamefont {Klinovaja},\ and\ \citenamefont {Loss}}]{mook2021interaction}%
  \BibitemOpen
  \bibfield  {author} {\bibinfo {author} {\bibfnamefont {A.}~\bibnamefont {Mook}}, \bibinfo {author} {\bibfnamefont {K.}~\bibnamefont {Plekhanov}}, \bibinfo {author} {\bibfnamefont {J.}~\bibnamefont {Klinovaja}}, \ and\ \bibinfo {author} {\bibfnamefont {D.}~\bibnamefont {Loss}},\ }\bibfield  {title} {\enquote {\bibinfo {title} {Interaction-stabilized topological magnon insulator in ferromagnets},}\ }\href@noop {} {\bibfield  {journal} {\bibinfo  {journal} {Physical Review X}\ }\textbf {\bibinfo {volume} {11}},\ \bibinfo {pages} {021061} (\bibinfo {year} {2021})}\BibitemShut {NoStop}%
\bibitem [{\citenamefont {Mazzamurro}\ \emph {et~al.}(2020)\citenamefont {Mazzamurro}, \citenamefont {Dusch}, \citenamefont {Pernod}, \citenamefont {Bou~Matar}, \citenamefont {Addad}, \citenamefont {Talbi},\ and\ \citenamefont {Tiercelin}}]{mazzamurro2020giant}%
  \BibitemOpen
  \bibfield  {author} {\bibinfo {author} {\bibfnamefont {A.}~\bibnamefont {Mazzamurro}}, \bibinfo {author} {\bibfnamefont {Y.}~\bibnamefont {Dusch}}, \bibinfo {author} {\bibfnamefont {P.}~\bibnamefont {Pernod}}, \bibinfo {author} {\bibfnamefont {O.}~\bibnamefont {Bou~Matar}}, \bibinfo {author} {\bibfnamefont {A.}~\bibnamefont {Addad}}, \bibinfo {author} {\bibfnamefont {A.}~\bibnamefont {Talbi}}, \ and\ \bibinfo {author} {\bibfnamefont {N.}~\bibnamefont {Tiercelin}},\ }\bibfield  {title} {\enquote {\bibinfo {title} {Giant magnetoelastic coupling in a love acoustic waveguide based on tb co 2/fe co nanostructured film on st-cut quartz},}\ }\href@noop {} {\bibfield  {journal} {\bibinfo  {journal} {Physical Review Applied}\ }\textbf {\bibinfo {volume} {13}},\ \bibinfo {pages} {044001} (\bibinfo {year} {2020})}\BibitemShut {NoStop}%
\bibitem [{\citenamefont {Okamoto}, \citenamefont {Murakami},\ and\ \citenamefont {Everschor-Sitte}(2020)}]{okamoto2020berry}%
  \BibitemOpen
  \bibfield  {author} {\bibinfo {author} {\bibfnamefont {A.}~\bibnamefont {Okamoto}}, \bibinfo {author} {\bibfnamefont {S.}~\bibnamefont {Murakami}}, \ and\ \bibinfo {author} {\bibfnamefont {K.}~\bibnamefont {Everschor-Sitte}},\ }\bibfield  {title} {\enquote {\bibinfo {title} {Berry curvature for magnetoelastic waves},}\ }\href@noop {} {\bibfield  {journal} {\bibinfo  {journal} {Physical Review B}\ }\textbf {\bibinfo {volume} {101}},\ \bibinfo {pages} {064424} (\bibinfo {year} {2020})}\BibitemShut {NoStop}%
\bibitem [{\citenamefont {Vonsovski{\u\i}}(1974)}]{vonsovskiĭ1974magnetism}%
  \BibitemOpen
  \bibfield  {author} {\bibinfo {author} {\bibfnamefont {S.}~\bibnamefont {Vonsovski{\u\i}}},\ }\href {https://books.google.ru/books?id=grzvAAAAMAAJ} {\emph {\bibinfo {title} {Magnetism:}}},\ \bibinfo {series} {Halsted Press Book}\ No.~\bibinfo {number} {1}\ (\bibinfo  {publisher} {J. Wiley},\ \bibinfo {year} {1974})\BibitemShut {NoStop}%
\bibitem [{\citenamefont {Turov}\ \emph {et~al.}(2001)\citenamefont {Turov}, \citenamefont {Kolchanov}, \citenamefont {Men’Shenin}, \citenamefont {Mirsaev},\ and\ \citenamefont {Nikolaev}}]{turov2001symmetry}%
  \BibitemOpen
  \bibfield  {author} {\bibinfo {author} {\bibfnamefont {E.}~\bibnamefont {Turov}}, \bibinfo {author} {\bibfnamefont {A.}~\bibnamefont {Kolchanov}}, \bibinfo {author} {\bibfnamefont {V.}~\bibnamefont {Men’Shenin}}, \bibinfo {author} {\bibfnamefont {I.}~\bibnamefont {Mirsaev}}, \ and\ \bibinfo {author} {\bibfnamefont {V.}~\bibnamefont {Nikolaev}},\ }\bibfield  {title} {\enquote {\bibinfo {title} {Symmetry and physical properties of antiferromagnets},}\ }\href@noop {} {\bibfield  {journal} {\bibinfo  {journal} {Fizmatlit, Moskow}\ } (\bibinfo {year} {2001})}\BibitemShut {NoStop}%
\bibitem [{\citenamefont {Zvezdin}\ and\ \citenamefont {Pyatakov}(2004)}]{zvezdin2004phase}%
  \BibitemOpen
  \bibfield  {author} {\bibinfo {author} {\bibfnamefont {A.~K.}\ \bibnamefont {Zvezdin}}\ and\ \bibinfo {author} {\bibfnamefont {A.~P.}\ \bibnamefont {Pyatakov}},\ }\bibfield  {title} {\enquote {\bibinfo {title} {Phase transitions and the giant magnetoelectric effect in multiferroics},}\ }\href@noop {} {\bibfield  {journal} {\bibinfo  {journal} {Physics-Uspekhi}\ }\textbf {\bibinfo {volume} {47}},\ \bibinfo {pages} {416} (\bibinfo {year} {2004})}\BibitemShut {NoStop}%
\bibitem [{\citenamefont {Turov}\ and\ \citenamefont {Nikolaev}(2005)}]{turov2005new}%
  \BibitemOpen
  \bibfield  {author} {\bibinfo {author} {\bibfnamefont {E.~A.}\ \bibnamefont {Turov}}\ and\ \bibinfo {author} {\bibfnamefont {V.~V.}\ \bibnamefont {Nikolaev}},\ }\bibfield  {title} {\enquote {\bibinfo {title} {New physical phenomena caused by magnetoelectric and antiferroelectric interactions in magnets},}\ }\href@noop {} {\bibfield  {journal} {\bibinfo  {journal} {Physics-Uspekhi}\ }\textbf {\bibinfo {volume} {48}},\ \bibinfo {pages} {431} (\bibinfo {year} {2005})}\BibitemShut {NoStop}%
\bibitem [{\citenamefont {Nagaev}(1988)}]{nagaev1988magnets}%
  \BibitemOpen
  \bibfield  {author} {\bibinfo {author} {\bibfnamefont {E.}~\bibnamefont {Nagaev}},\ }\href@noop {} {\enquote {\bibinfo {title} {Magnets with complex exchange interactions},}\ } (\bibinfo {year} {1988})\BibitemShut {NoStop}%
\bibitem [{\citenamefont {Patashinskii}\ and\ \citenamefont {Pokrovskii}(1979)}]{patashinskii1979fluctuation}%
  \BibitemOpen
  \bibfield  {author} {\bibinfo {author} {\bibfnamefont {A.}~\bibnamefont {Patashinskii}}\ and\ \bibinfo {author} {\bibfnamefont {V.}~\bibnamefont {Pokrovskii}},\ }\href@noop {} {\emph {\bibinfo {title} {Fluctuation theory of phase transitions}}}\ (\bibinfo  {publisher} {Oxford, New York},\ \bibinfo {year} {1979})\BibitemShut {NoStop}%
\bibitem [{\citenamefont {Dikshtein}, \citenamefont {Tarasenko},\ and\ \citenamefont {Sharvov}(1974)}]{dikshtein1974effect}%
  \BibitemOpen
  \bibfield  {author} {\bibinfo {author} {\bibfnamefont {I.}~\bibnamefont {Dikshtein}}, \bibinfo {author} {\bibfnamefont {V.}~\bibnamefont {Tarasenko}}, \ and\ \bibinfo {author} {\bibfnamefont {V.}~\bibnamefont {Sharvov}},\ }\bibfield  {title} {\enquote {\bibinfo {title} {Effect of pressure on the resonance properties of uniaxial ferro- and antiferromagnetics},}\ }\href@noop {} {\bibfield  {journal} {\bibinfo  {journal} {Fizika Tverdogo Tela}\ }\textbf {\bibinfo {volume} {16}},\ \bibinfo {pages} {2192--2197} (\bibinfo {year} {1974})}\BibitemShut {NoStop}%
\bibitem [{\citenamefont {Dikshtein}, \citenamefont {Tarasenko},\ and\ \citenamefont {Shavrov}(1974)}]{dikshtein1974effects}%
  \BibitemOpen
  \bibfield  {author} {\bibinfo {author} {\bibfnamefont {I.}~\bibnamefont {Dikshtein}}, \bibinfo {author} {\bibfnamefont {V.}~\bibnamefont {Tarasenko}}, \ and\ \bibinfo {author} {\bibfnamefont {V.}~\bibnamefont {Shavrov}},\ }\bibfield  {title} {\enquote {\bibinfo {title} {Effect of pressure on magnetoacoustic resonance in uniaxial antiferromagnets},}\ }\href@noop {} {\bibfield  {journal} {\bibinfo  {journal} {Zh. eksp. teor. fiz}\ }\textbf {\bibinfo {volume} {67}},\ \bibinfo {pages} {816--823} (\bibinfo {year} {1974})}\BibitemShut {NoStop}%
\bibitem [{\citenamefont {Turov}\ and\ \citenamefont {Shavrov}(1983)}]{turov1983broken}%
  \BibitemOpen
  \bibfield  {author} {\bibinfo {author} {\bibfnamefont {E.~A.}\ \bibnamefont {Turov}}\ and\ \bibinfo {author} {\bibfnamefont {V.~G.}\ \bibnamefont {Shavrov}},\ }\bibfield  {title} {\enquote {\bibinfo {title} {Broken symmetry and magnetoacoustic effects in ferroand antiferromagnetics},}\ }\href@noop {} {\bibfield  {journal} {\bibinfo  {journal} {Soviet Physics Uspekhi}\ }\textbf {\bibinfo {volume} {26}},\ \bibinfo {pages} {593} (\bibinfo {year} {1983})}\BibitemShut {NoStop}%
\bibitem [{\citenamefont {Bar'yakhtar}\ \emph {et~al.}(1984)\citenamefont {Bar'yakhtar}, \citenamefont {Vitebskii}, \citenamefont {Pashkevich}, \citenamefont {Sobolev},\ and\ \citenamefont {Tarasenko}}]{bar1984striction}%
  \BibitemOpen
  \bibfield  {author} {\bibinfo {author} {\bibfnamefont {V.}~\bibnamefont {Bar'yakhtar}}, \bibinfo {author} {\bibfnamefont {I.}~\bibnamefont {Vitebskii}}, \bibinfo {author} {\bibfnamefont {Y.~G.}\ \bibnamefont {Pashkevich}}, \bibinfo {author} {\bibfnamefont {V.}~\bibnamefont {Sobolev}}, \ and\ \bibinfo {author} {\bibfnamefont {V.}~\bibnamefont {Tarasenko}},\ }\bibfield  {title} {\enquote {\bibinfo {title} {Striction effects and dynamics of the magnetic subsystem in spin-reorientation phase transitions. symmetry aspects},}\ }\href@noop {} {\bibfield  {journal} {\bibinfo  {journal} {Zh. Eksp. Teor. Fiz}\ }\textbf {\bibinfo {volume} {87}},\ \bibinfo {pages} {1028--1037} (\bibinfo {year} {1984})}\BibitemShut {NoStop}%
\bibitem [{\citenamefont {Buchel'nikov}\ \emph {et~al.}(1996)\citenamefont {Buchel'nikov}, \citenamefont {Dan'shin}, \citenamefont {Tsymbal},\ and\ \citenamefont {Shavrov}}]{buchel1996magnetoacoustics}%
  \BibitemOpen
  \bibfield  {author} {\bibinfo {author} {\bibfnamefont {V.~D.}\ \bibnamefont {Buchel'nikov}}, \bibinfo {author} {\bibfnamefont {N.~K.}\ \bibnamefont {Dan'shin}}, \bibinfo {author} {\bibfnamefont {L.}~\bibnamefont {Tsymbal}}, \ and\ \bibinfo {author} {\bibfnamefont {V.~G.}\ \bibnamefont {Shavrov}},\ }\bibfield  {title} {\enquote {\bibinfo {title} {Magnetoacoustics of rare-earth orthoferrites},}\ }\href@noop {} {\bibfield  {journal} {\bibinfo  {journal} {Physics-Uspekhi}\ }\textbf {\bibinfo {volume} {39}},\ \bibinfo {pages} {547} (\bibinfo {year} {1996})}\BibitemShut {NoStop}%
\bibitem [{\citenamefont {Buchel'nikov}\ \emph {et~al.}(1999)\citenamefont {Buchel'nikov}, \citenamefont {Dan'shin}, \citenamefont {Tsymbal},\ and\ \citenamefont {Shavrov}}]{buchel1999relative}%
  \BibitemOpen
  \bibfield  {author} {\bibinfo {author} {\bibfnamefont {V.~D.}\ \bibnamefont {Buchel'nikov}}, \bibinfo {author} {\bibfnamefont {N.~K.}\ \bibnamefont {Dan'shin}}, \bibinfo {author} {\bibfnamefont {L.}~\bibnamefont {Tsymbal}}, \ and\ \bibinfo {author} {\bibfnamefont {V.~G.}\ \bibnamefont {Shavrov}},\ }\bibfield  {title} {\enquote {\bibinfo {title} {On the relative contributions of precessional and longitudinal oscillations to the dynamics of magnets},}\ }\href@noop {} {\bibfield  {journal} {\bibinfo  {journal} {Physics-Uspekhi}\ }\textbf {\bibinfo {volume} {42}},\ \bibinfo {pages} {957} (\bibinfo {year} {1999})}\BibitemShut {NoStop}%
\bibitem [{\citenamefont {Turov}\ and\ \citenamefont {Irkhin}(1956)}]{turov1956vibration}%
  \BibitemOpen
  \bibfield  {author} {\bibinfo {author} {\bibfnamefont {E.}~\bibnamefont {Turov}}\ and\ \bibinfo {author} {\bibfnamefont {Y.~P.}\ \bibnamefont {Irkhin}},\ }\bibfield  {title} {\enquote {\bibinfo {title} {On the vibration spectrum of ferromagnetic elastic medium},}\ }\href@noop {} {\bibfield  {journal} {\bibinfo  {journal} {Fiz. Met. Metalloved.}\ }\textbf {\bibinfo {volume} {3}},\ \bibinfo {pages} {15--17} (\bibinfo {year} {1956})}\BibitemShut {NoStop}%
\bibitem [{\citenamefont {Akhiezer}, \citenamefont {Baryakhtar},\ and\ \citenamefont {Peletminsky}(1958)}]{akhiezer1958coupled}%
  \BibitemOpen
  \bibfield  {author} {\bibinfo {author} {\bibfnamefont {A.}~\bibnamefont {Akhiezer}}, \bibinfo {author} {\bibfnamefont {V.}~\bibnamefont {Baryakhtar}}, \ and\ \bibinfo {author} {\bibfnamefont {S.}~\bibnamefont {Peletminsky}},\ }\bibfield  {title} {\enquote {\bibinfo {title} {Coupled magnetoelastic waves in ferromagnetic and ferracoustic resonance},}\ }\href@noop {} {\bibfield  {journal} {\bibinfo  {journal} {Zh. Exper. Teor. Fiz}\ }\textbf {\bibinfo {volume} {35}},\ \bibinfo {pages} {228--236} (\bibinfo {year} {1958})}\BibitemShut {NoStop}%
\bibitem [{\citenamefont {Kittel}(1958)}]{kittel1958interaction}%
  \BibitemOpen
  \bibfield  {author} {\bibinfo {author} {\bibfnamefont {C.}~\bibnamefont {Kittel}},\ }\bibfield  {title} {\enquote {\bibinfo {title} {Interaction of spin waves and ultrasonic waves in ferromagnetic crystals},}\ }\href@noop {} {\bibfield  {journal} {\bibinfo  {journal} {Physical Review}\ }\textbf {\bibinfo {volume} {110}},\ \bibinfo {pages} {836} (\bibinfo {year} {1958})}\BibitemShut {NoStop}%
\bibitem [{\citenamefont {Peletminskii}(1960)}]{peletminskii1960coupled}%
  \BibitemOpen
  \bibfield  {author} {\bibinfo {author} {\bibfnamefont {S.}~\bibnamefont {Peletminskii}},\ }\bibfield  {title} {\enquote {\bibinfo {title} {Coupled magnetoelastic oscillations in antiferromagnetics},}\ }\href@noop {} {\bibfield  {journal} {\bibinfo  {journal} {Soviet Physics JETP}\ }\textbf {\bibinfo {volume} {37}},\ \bibinfo {pages} {321--324} (\bibinfo {year} {1960})}\BibitemShut {NoStop}%
\bibitem [{\citenamefont {Belov}(1987)}]{belov1987magnetostriction}%
  \BibitemOpen
  \bibfield  {author} {\bibinfo {author} {\bibfnamefont {K.}~\bibnamefont {Belov}},\ }\bibfield  {title} {\enquote {\bibinfo {title} {Magnetostriction phenomena and their technical applications},}\ }\href@noop {} {\bibfield  {journal} {\bibinfo  {journal} {M.: Science}\ ,\ \bibinfo {pages} {160}} (\bibinfo {year} {1987})}\BibitemShut {NoStop}%
\bibitem [{\citenamefont {Goldin}\ \emph {et~al.}(1991)\citenamefont {Goldin}, \citenamefont {Kotov}, \citenamefont {Zarembo},\ and\ \citenamefont {Karpachev}}]{goldin1991spin}%
  \BibitemOpen
  \bibfield  {author} {\bibinfo {author} {\bibfnamefont {B.}~\bibnamefont {Goldin}}, \bibinfo {author} {\bibfnamefont {L.}~\bibnamefont {Kotov}}, \bibinfo {author} {\bibfnamefont {L.}~\bibnamefont {Zarembo}}, \ and\ \bibinfo {author} {\bibfnamefont {S.}~\bibnamefont {Karpachev}},\ }\href@noop {} {\enquote {\bibinfo {title} {Spin-phonon interactions in crystals (ferrites) [in russian]},}\ } (\bibinfo {year} {1991})\BibitemShut {NoStop}%
\bibitem [{\citenamefont {Tacker}\ and\ \citenamefont {Rampton}(1975)}]{tacker1975hypersound}%
  \BibitemOpen
  \bibfield  {author} {\bibinfo {author} {\bibfnamefont {J.}~\bibnamefont {Tacker}}\ and\ \bibinfo {author} {\bibfnamefont {V.}~\bibnamefont {Rampton}},\ }\href@noop {} {\enquote {\bibinfo {title} {Hypersound with applications in solid-state physics [russian translation]},}\ } (\bibinfo {year} {1975})\BibitemShut {NoStop}%
\bibitem [{\citenamefont {Tasaki}\ \emph {et~al.}(1963)\citenamefont {Tasaki} \emph {et~al.}}]{tasaki1963magnetic}%
  \BibitemOpen
  \bibfield  {author} {\bibinfo {author} {\bibfnamefont {A.}~\bibnamefont {Tasaki}} \emph {et~al.},\ }\bibfield  {title} {\enquote {\bibinfo {title} {Magnetic properties of synthetic single crystal of $\alpha$fe2o3},}\ }\href@noop {} {\bibfield  {journal} {\bibinfo  {journal} {Journal of the Physical Society of Japan}\ }\textbf {\bibinfo {volume} {18}},\ \bibinfo {pages} {1148--1154} (\bibinfo {year} {1963})}\BibitemShut {NoStop}%
\bibitem [{\citenamefont {Mizushima}\ and\ \citenamefont {Iida}(1966)}]{mizushima1966effective}%
  \BibitemOpen
  \bibfield  {author} {\bibinfo {author} {\bibfnamefont {K.}~\bibnamefont {Mizushima}}\ and\ \bibinfo {author} {\bibfnamefont {S.}~\bibnamefont {Iida}},\ }\bibfield  {title} {\enquote {\bibinfo {title} {Effective in-plane anisotropy field in $\alpha$fe2o3},}\ }\href@noop {} {\bibfield  {journal} {\bibinfo  {journal} {Journal of the Physical Society of Japan}\ }\textbf {\bibinfo {volume} {21}},\ \bibinfo {pages} {1521--1526} (\bibinfo {year} {1966})}\BibitemShut {NoStop}%
\bibitem [{\citenamefont {Iida}\ and\ \citenamefont {Tasaki}(1964)}]{iida1964magnetoelastic}%
  \BibitemOpen
  \bibfield  {author} {\bibinfo {author} {\bibfnamefont {S.}~\bibnamefont {Iida}}\ and\ \bibinfo {author} {\bibfnamefont {A.}~\bibnamefont {Tasaki}},\ }\bibfield  {title} {\enquote {\bibinfo {title} {Magnetoelastic coupling in parasitic ferromagnet $\alpha$-fe2o3},}\ }in\ \href@noop {} {\emph {\bibinfo {booktitle} {Proc. Int. Conf. Magn. Nottingham}}}\ (\bibinfo {year} {1964})\ p.\ \bibinfo {pages} {583}\BibitemShut {NoStop}%
\bibitem [{\citenamefont {Cooper}(1968{\natexlab{a}})}]{cooper1968spin}%
  \BibitemOpen
  \bibfield  {author} {\bibinfo {author} {\bibfnamefont {B.~R.}\ \bibnamefont {Cooper}},\ }\bibfield  {title} {\enquote {\bibinfo {title} {Spin waves and magnetic resonance in rare-earth metals: thermal, applied-field, and magnetoelastic effects},}\ }\href@noop {} {\bibfield  {journal} {\bibinfo  {journal} {Physical Review}\ }\textbf {\bibinfo {volume} {169}},\ \bibinfo {pages} {281} (\bibinfo {year} {1968}{\natexlab{a}})}\BibitemShut {NoStop}%
\bibitem [{\citenamefont {Cooper}(1968{\natexlab{b}})}]{cooper1968magnetic}%
  \BibitemOpen
  \bibfield  {author} {\bibinfo {author} {\bibfnamefont {B.~R.}\ \bibnamefont {Cooper}},\ }\bibfield  {title} {\enquote {\bibinfo {title} {Magnetic properties of rare earth metals},}\ }in\ \href@noop {} {\emph {\bibinfo {booktitle} {Solid State Physics}}},\ Vol.~\bibinfo {volume} {21}\ (\bibinfo  {publisher} {Elsevier},\ \bibinfo {year} {1968})\ pp.\ \bibinfo {pages} {393--490}\BibitemShut {NoStop}%
\bibitem [{\citenamefont {Cooper}(1972)}]{cooper1972phenomenological}%
  \BibitemOpen
  \bibfield  {author} {\bibinfo {author} {\bibfnamefont {B.~R.}\ \bibnamefont {Cooper}},\ }\bibfield  {title} {\enquote {\bibinfo {title} {Phenomenological theory of magnetic ordering: Importance of interactions with the crystal lattice},}\ }in\ \href@noop {} {\emph {\bibinfo {booktitle} {Magnetic properties of rare earth metals}}}\ (\bibinfo  {publisher} {Springer},\ \bibinfo {year} {1972})\ pp.\ \bibinfo {pages} {17--80}\BibitemShut {NoStop}%
\bibitem [{\citenamefont {Nielsen}\ \emph {et~al.}(1970)\citenamefont {Nielsen}, \citenamefont {M{\o}ller}, \citenamefont {Lindg{\aa}rd},\ and\ \citenamefont {Mackintosh}}]{nielsen1970magnetic}%
  \BibitemOpen
  \bibfield  {author} {\bibinfo {author} {\bibfnamefont {M.}~\bibnamefont {Nielsen}}, \bibinfo {author} {\bibfnamefont {H.~B.}\ \bibnamefont {M{\o}ller}}, \bibinfo {author} {\bibfnamefont {P.-A.}\ \bibnamefont {Lindg{\aa}rd}}, \ and\ \bibinfo {author} {\bibfnamefont {A.}~\bibnamefont {Mackintosh}},\ }\bibfield  {title} {\enquote {\bibinfo {title} {Magnetic anisotropy in rare-earth metals},}\ }\href@noop {} {\bibfield  {journal} {\bibinfo  {journal} {Physical Review Letters}\ }\textbf {\bibinfo {volume} {25}},\ \bibinfo {pages} {1451} (\bibinfo {year} {1970})}\BibitemShut {NoStop}%
\bibitem [{\citenamefont {Nielsen}, \citenamefont {M{\o}ller},\ and\ \citenamefont {Mackintosh}(1970)}]{nielsen1970magnon}%
  \BibitemOpen
  \bibfield  {author} {\bibinfo {author} {\bibfnamefont {M.}~\bibnamefont {Nielsen}}, \bibinfo {author} {\bibfnamefont {H.~B.}\ \bibnamefont {M{\o}ller}}, \ and\ \bibinfo {author} {\bibfnamefont {A.}~\bibnamefont {Mackintosh}},\ }\bibfield  {title} {\enquote {\bibinfo {title} {Magnon interactions in terbium},}\ }\href@noop {} {\bibfield  {journal} {\bibinfo  {journal} {Journal of Applied Physics}\ }\textbf {\bibinfo {volume} {41}},\ \bibinfo {pages} {1174--1175} (\bibinfo {year} {1970})}\BibitemShut {NoStop}%
\bibitem [{\citenamefont {Ozhogin}\ and\ \citenamefont {Maximenkov}(1972)}]{ozhogin1972easy}%
  \BibitemOpen
  \bibfield  {author} {\bibinfo {author} {\bibfnamefont {V.}~\bibnamefont {Ozhogin}}\ and\ \bibinfo {author} {\bibfnamefont {P.}~\bibnamefont {Maximenkov}},\ }\bibfield  {title} {\enquote {\bibinfo {title} {Easy plane antiferromagnets (afep) for applications: Hematite},}\ }\href@noop {} {\bibfield  {journal} {\bibinfo  {journal} {IEEE Transactions on Magnetics}\ }\textbf {\bibinfo {volume} {8}},\ \bibinfo {pages} {645--645} (\bibinfo {year} {1972})}\BibitemShut {NoStop}%
\bibitem [{\citenamefont {Seavey}(1972)}]{seavey1972acoustic}%
  \BibitemOpen
  \bibfield  {author} {\bibinfo {author} {\bibfnamefont {M.}~\bibnamefont {Seavey}},\ }\bibfield  {title} {\enquote {\bibinfo {title} {Acoustic resonance in the easy-plane weak ferromagnets alpha;- fe2o3 and febo3},}\ }\href@noop {} {\bibfield  {journal} {\bibinfo  {journal} {Solid State Communications}\ }\textbf {\bibinfo {volume} {10}},\ \bibinfo {pages} {219--223} (\bibinfo {year} {1972})}\BibitemShut {NoStop}%
\bibitem [{\citenamefont {Maksimenkov}\ and\ \citenamefont {Ozhogin}(1973)}]{maksimenkov1973antiferromagnetic}%
  \BibitemOpen
  \bibfield  {author} {\bibinfo {author} {\bibfnamefont {P.}~\bibnamefont {Maksimenkov}}\ and\ \bibinfo {author} {\bibfnamefont {V.}~\bibnamefont {Ozhogin}},\ }\bibfield  {title} {\enquote {\bibinfo {title} {Antiferromagnetic resonance study of the magnetoelastic interactions in hematite},}\ }\href@noop {} {\bibfield  {journal} {\bibinfo  {journal} {Zhurnal Eksperimental'noi i Teoreticheskoi Fiziki}\ }\textbf {\bibinfo {volume} {65}},\ \bibinfo {pages} {657--67} (\bibinfo {year} {1973})}\BibitemShut {NoStop}%
\bibitem [{\citenamefont {Tantz}\ and\ \citenamefont {Wetting}(1978)}]{tantz1978spin}%
  \BibitemOpen
  \bibfield  {author} {\bibinfo {author} {\bibfnamefont {W.}~\bibnamefont {Tantz}}\ and\ \bibinfo {author} {\bibfnamefont {W.}~\bibnamefont {Wetting}},\ }\bibfield  {title} {\enquote {\bibinfo {title} {Spin wave dispersion of febo3 at small wavevector},}\ }\href@noop {} {\bibfield  {journal} {\bibinfo  {journal} {J. Appl. Phys}\ }\textbf {\bibinfo {volume} {15}},\ \bibinfo {pages} {399} (\bibinfo {year} {1978})}\BibitemShut {NoStop}%
\bibitem [{\citenamefont {Gorodetsky}\ and\ \citenamefont {L{\"u}thi}(1970)}]{gorodetsky1970sound}%
  \BibitemOpen
  \bibfield  {author} {\bibinfo {author} {\bibfnamefont {G.}~\bibnamefont {Gorodetsky}}\ and\ \bibinfo {author} {\bibfnamefont {B.}~\bibnamefont {L{\"u}thi}},\ }\bibfield  {title} {\enquote {\bibinfo {title} {Sound-wave—soft-mode interaction near displacive phase transitions: Spin reorientation in erfe o 3},}\ }\href@noop {} {\bibfield  {journal} {\bibinfo  {journal} {Physical Review B}\ }\textbf {\bibinfo {volume} {2}},\ \bibinfo {pages} {3688} (\bibinfo {year} {1970})}\BibitemShut {NoStop}%
\bibitem [{\citenamefont {Grishmanovskij}\ \emph {et~al.}(1974)\citenamefont {Grishmanovskij}, \citenamefont {Lemanov}, \citenamefont {Smolenskij}, \citenamefont {Balbashov},\ and\ \citenamefont {Chervonenkis}}]{grishmanovskij1974piezomagnetic}%
  \BibitemOpen
  \bibfield  {author} {\bibinfo {author} {\bibfnamefont {A.}~\bibnamefont {Grishmanovskij}}, \bibinfo {author} {\bibfnamefont {V.}~\bibnamefont {Lemanov}}, \bibinfo {author} {\bibfnamefont {G.}~\bibnamefont {Smolenskij}}, \bibinfo {author} {\bibfnamefont {A.}~\bibnamefont {Balbashov}}, \ and\ \bibinfo {author} {\bibfnamefont {A.~Y.}\ \bibnamefont {Chervonenkis}},\ }\bibfield  {title} {\enquote {\bibinfo {title} {Piezomagnetic and magnetoelastic effects during the propagation of elastic waves in crystals of rare earth orthoferrites},}\ }\href@noop {} {\bibfield  {journal} {\bibinfo  {journal} {Fizika Tverdogo Tela}\ }\textbf {\bibinfo {volume} {16}},\ \bibinfo {pages} {1426--1431} (\bibinfo {year} {1974})}\BibitemShut {NoStop}%
\bibitem [{\citenamefont {Gorodetsky}, \citenamefont {Shaft},\ and\ \citenamefont {Wanklyn}(1976)}]{gorodetsky1976magnetoelastic}%
  \BibitemOpen
  \bibfield  {author} {\bibinfo {author} {\bibfnamefont {G.}~\bibnamefont {Gorodetsky}}, \bibinfo {author} {\bibfnamefont {S.}~\bibnamefont {Shaft}}, \ and\ \bibinfo {author} {\bibfnamefont {B.}~\bibnamefont {Wanklyn}},\ }\bibfield  {title} {\enquote {\bibinfo {title} {Magnetoelastic properties of tmfe o 3 at the spin reorientation region},}\ }\href@noop {} {\bibfield  {journal} {\bibinfo  {journal} {Physical Review B}\ }\textbf {\bibinfo {volume} {14}},\ \bibinfo {pages} {2051} (\bibinfo {year} {1976})}\BibitemShut {NoStop}%
\bibitem [{\citenamefont {Shapira}(1967)}]{shapira1967absorption}%
  \BibitemOpen
  \bibfield  {author} {\bibinfo {author} {\bibfnamefont {Y.}~\bibnamefont {Shapira}},\ }\bibfield  {title} {\enquote {\bibinfo {title} {Absorption peak for ultrasonic waves near the spin-flop transition of uniaxial antiferromagnets},}\ }\href@noop {} {\bibfield  {journal} {\bibinfo  {journal} {Physics Letters A}\ }\textbf {\bibinfo {volume} {24}},\ \bibinfo {pages} {361--362} (\bibinfo {year} {1967})}\BibitemShut {NoStop}%
\bibitem [{\citenamefont {Shapira}\ and\ \citenamefont {Zak}(1968)}]{shapira1968ultrasonic}%
  \BibitemOpen
  \bibfield  {author} {\bibinfo {author} {\bibfnamefont {Y.}~\bibnamefont {Shapira}}\ and\ \bibinfo {author} {\bibfnamefont {J.}~\bibnamefont {Zak}},\ }\bibfield  {title} {\enquote {\bibinfo {title} {Ultrasonic attenuation near and above the spin-flop transition of mn f 2},}\ }\href@noop {} {\bibfield  {journal} {\bibinfo  {journal} {Physical Review}\ }\textbf {\bibinfo {volume} {170}},\ \bibinfo {pages} {503} (\bibinfo {year} {1968})}\BibitemShut {NoStop}%
\bibitem [{\citenamefont {Shapira}(1969)}]{shapira1969ultrasonic}%
  \BibitemOpen
  \bibfield  {author} {\bibinfo {author} {\bibfnamefont {Y.}~\bibnamefont {Shapira}},\ }\bibfield  {title} {\enquote {\bibinfo {title} {Ultrasonic behavior near the spin-flop transition of cr 2 o 3},}\ }\href@noop {} {\bibfield  {journal} {\bibinfo  {journal} {Physical Review}\ }\textbf {\bibinfo {volume} {187}},\ \bibinfo {pages} {734} (\bibinfo {year} {1969})}\BibitemShut {NoStop}%
\bibitem [{\citenamefont {Chepurnykh}(1975{\natexlab{a}})}]{chepurnykh1975some}%
  \BibitemOpen
  \bibfield  {author} {\bibinfo {author} {\bibfnamefont {G.}~\bibnamefont {Chepurnykh}},\ }\bibfield  {title} {\enquote {\bibinfo {title} {Some pecularieties of interaction of magnetoelastic waves in uniaxial antiferromagnets},}\ }\href@noop {} {\bibfield  {journal} {\bibinfo  {journal} {Fizika Tverdogo Tela}\ }\textbf {\bibinfo {volume} {17}},\ \bibinfo {pages} {430--432} (\bibinfo {year} {1975}{\natexlab{a}})}\BibitemShut {NoStop}%
\bibitem [{\citenamefont {Chepurnykh}(1975{\natexlab{b}})}]{chepurnykh1975magnetostriction}%
  \BibitemOpen
  \bibfield  {author} {\bibinfo {author} {\bibfnamefont {G.}~\bibnamefont {Chepurnykh}},\ }\href@noop {} {\enquote {\bibinfo {title} {Magnetostriction effect on overturing of sublattices of antiferromagnet},}\ } (\bibinfo {year} {1975}{\natexlab{b}})\BibitemShut {NoStop}%
\bibitem [{\citenamefont {Dikshtein}, \citenamefont {Tarasenko},\ and\ \citenamefont {Shavrov}(1976)}]{dikshtein1976spin}%
  \BibitemOpen
  \bibfield  {author} {\bibinfo {author} {\bibfnamefont {I.}~\bibnamefont {Dikshtein}}, \bibinfo {author} {\bibfnamefont {V.}~\bibnamefont {Tarasenko}}, \ and\ \bibinfo {author} {\bibfnamefont {V.}~\bibnamefont {Shavrov}},\ }\bibfield  {title} {\enquote {\bibinfo {title} {Spin-spin resonance in antiferromagnetics under pressure},}\ }\href@noop {} {\bibfield  {journal} {\bibinfo  {journal} {Fiz. Met. Metalloved.}\ }\textbf {\bibinfo {volume} {42}},\ \bibinfo {pages} {903--907} (\bibinfo {year} {1976})}\BibitemShut {NoStop}%
\bibitem [{\citenamefont {Buchel'nikov}, \citenamefont {Tarasenko},\ and\ \citenamefont {Shavrov}(1983)}]{buchel1983relaxation}%
  \BibitemOpen
  \bibfield  {author} {\bibinfo {author} {\bibfnamefont {V.~D.}\ \bibnamefont {Buchel'nikov}}, \bibinfo {author} {\bibfnamefont {V.}~\bibnamefont {Tarasenko}}, \ and\ \bibinfo {author} {\bibfnamefont {V.~G.}\ \bibnamefont {Shavrov}},\ }\bibfield  {title} {\enquote {\bibinfo {title} {Relaxation processes in magnetics near orientational phase transitions},}\ }\href@noop {} {\bibfield  {journal} {\bibinfo  {journal} {Fizika Tverdogo Tela}\ }\textbf {\bibinfo {volume} {25}},\ \bibinfo {pages} {3019--3024} (\bibinfo {year} {1983})}\BibitemShut {NoStop}%
\bibitem [{\citenamefont {Buchel'nikov}\ and\ \citenamefont {Shavrov}(1984)}]{buchel1984magnon}%
  \BibitemOpen
  \bibfield  {author} {\bibinfo {author} {\bibfnamefont {V.~D.}\ \bibnamefont {Buchel'nikov}}\ and\ \bibinfo {author} {\bibfnamefont {V.~G.}\ \bibnamefont {Shavrov}},\ }\bibfield  {title} {\enquote {\bibinfo {title} {Magnon attenuation in antiferromagnets near orientational phase transition},}\ }\href@noop {} {\bibfield  {journal} {\bibinfo  {journal} {Fizika Tverdogo Tela}\ }\textbf {\bibinfo {volume} {26}},\ \bibinfo {pages} {1254--1256} (\bibinfo {year} {1984})}\BibitemShut {NoStop}%
\bibitem [{\citenamefont {Ozhogin}\ and\ \citenamefont {Preobrazhenskii}(1977)}]{ozhogin1977effective}%
  \BibitemOpen
  \bibfield  {author} {\bibinfo {author} {\bibfnamefont {V.}~\bibnamefont {Ozhogin}}\ and\ \bibinfo {author} {\bibfnamefont {V.}~\bibnamefont {Preobrazhenskii}},\ }\bibfield  {title} {\enquote {\bibinfo {title} {Effective anharmonicity of elastic subsystem in antiferromagnets},}\ }\href@noop {} {\bibfield  {journal} {\bibinfo  {journal} {Physica B+ C}\ }\textbf {\bibinfo {volume} {86}},\ \bibinfo {pages} {979--981} (\bibinfo {year} {1977})}\BibitemShut {NoStop}%
\bibitem [{\citenamefont {Berezin}\ and\ \citenamefont {Shavrov}(1977)}]{berezin1977antiferromagnetic}%
  \BibitemOpen
  \bibfield  {author} {\bibinfo {author} {\bibfnamefont {A.}~\bibnamefont {Berezin}}\ and\ \bibinfo {author} {\bibfnamefont {V.}~\bibnamefont {Shavrov}},\ }\bibfield  {title} {\enquote {\bibinfo {title} {Antiferromagnetic resonance in cubic crystals},}\ }\href@noop {} {\bibfield  {journal} {\bibinfo  {journal} {Soviet Journal of Experimental and Theoretical Physics}\ }\textbf {\bibinfo {volume} {45}},\ \bibinfo {pages} {1242} (\bibinfo {year} {1977})}\BibitemShut {NoStop}%
\bibitem [{\citenamefont {Sokolov}\ and\ \citenamefont {Shevaleevskii}(1977)}]{sokolov1977antiferromagnetic}%
  \BibitemOpen
  \bibfield  {author} {\bibinfo {author} {\bibfnamefont {V.}~\bibnamefont {Sokolov}}\ and\ \bibinfo {author} {\bibfnamefont {O.}~\bibnamefont {Shevaleevskii}},\ }\bibfield  {title} {\enquote {\bibinfo {title} {Antiferromagnetic resonance in cubic crystals of fege and crge garnets},}\ }\href@noop {} {\bibfield  {journal} {\bibinfo  {journal} {Zhurnal Eksperimental'noi i Teoreticheskoi Fiziki}\ }\textbf {\bibinfo {volume} {72}},\ \bibinfo {pages} {2367--74} (\bibinfo {year} {1977})}\BibitemShut {NoStop}%
\bibitem [{\citenamefont {Buchel'nikov}\ and\ \citenamefont {Shavrov}(1983)}]{buchel1983magnetoacoustic}%
  \BibitemOpen
  \bibfield  {author} {\bibinfo {author} {\bibfnamefont {V.}~\bibnamefont {Buchel'nikov}}\ and\ \bibinfo {author} {\bibfnamefont {V.}~\bibnamefont {Shavrov}},\ }\bibfield  {title} {\enquote {\bibinfo {title} {Magnetoacoustic oscillations jn elastically strained cubic crystals},}\ }\href@noop {} {\bibfield  {journal} {\bibinfo  {journal} {Fizika Metallov i Metallovedenie}\ }\textbf {\bibinfo {volume} {55}},\ \bibinfo {pages} {892--900} (\bibinfo {year} {1983})}\BibitemShut {NoStop}%
\bibitem [{\citenamefont {Dan'shin}, \citenamefont {Zherlitsyn},\ and\ \citenamefont {Zvada}(1989)}]{dan1989dynamic}%
  \BibitemOpen
  \bibfield  {author} {\bibinfo {author} {\bibfnamefont {N.~K.}\ \bibnamefont {Dan'shin}}, \bibinfo {author} {\bibfnamefont {S.}~\bibnamefont {Zherlitsyn}}, \ and\ \bibinfo {author} {\bibfnamefont {S.}~\bibnamefont {Zvada}},\ }\bibfield  {title} {\enquote {\bibinfo {title} {Dynamic properties of hofeo 3 in the region of spin reorientation},}\ }\href@noop {} {\bibfield  {journal} {\bibinfo  {journal} {Fizika Tverdogo Tela}\ }\textbf {\bibinfo {volume} {31}},\ \bibinfo {pages} {198--204} (\bibinfo {year} {1989})}\BibitemShut {NoStop}%
\bibitem [{\citenamefont {Vitebskij}\ \emph {et~al.}(1990)\citenamefont {Vitebskij}, \citenamefont {Dan'shin}, \citenamefont {Izotov}, \citenamefont {Sdvizhkov},\ and\ \citenamefont {Tsymbal}}]{vitebskij1990anomalous}%
  \BibitemOpen
  \bibfield  {author} {\bibinfo {author} {\bibfnamefont {I.}~\bibnamefont {Vitebskij}}, \bibinfo {author} {\bibfnamefont {N.}~\bibnamefont {Dan'shin}}, \bibinfo {author} {\bibfnamefont {A.}~\bibnamefont {Izotov}}, \bibinfo {author} {\bibfnamefont {M.}~\bibnamefont {Sdvizhkov}}, \ and\ \bibinfo {author} {\bibfnamefont {L.}~\bibnamefont {Tsymbal}},\ }\bibfield  {title} {\enquote {\bibinfo {title} {Anomalous critical dynamics in the low-temperature transition in erbium orthoferrite},}\ }\href@noop {} {\bibfield  {journal} {\bibinfo  {journal} {Zhurnal Ehksperimental'noj i Teoreticheskoj Fiziki}\ }\textbf {\bibinfo {volume} {98}},\ \bibinfo {pages} {334--339} (\bibinfo {year} {1990})}\BibitemShut {NoStop}%
\bibitem [{\citenamefont {Buchelnikov}\ and\ \citenamefont {Shavrov}(1994)}]{buchelnikov1994influence}%
  \BibitemOpen
  \bibfield  {author} {\bibinfo {author} {\bibfnamefont {V.}~\bibnamefont {Buchelnikov}}\ and\ \bibinfo {author} {\bibfnamefont {V.}~\bibnamefont {Shavrov}},\ }\bibfield  {title} {\enquote {\bibinfo {title} {Effects of longitudinal susceptibility and relaxation on spin- and elastic-wave spectra in antiferrornagnets with of spin reorientation},}\ }\href@noop {} {\bibfield  {journal} {\bibinfo  {journal} {JETP}\ }\textbf {\bibinfo {volume} {79}},\ \bibinfo {pages} {951--955} (\bibinfo {year} {1994})}\BibitemShut {NoStop}%
\bibitem [{\citenamefont {Dan'shin}, \citenamefont {Nepochatykh},\ and\ \citenamefont {Shkar}(1996)}]{dan1996role}%
  \BibitemOpen
  \bibfield  {author} {\bibinfo {author} {\bibfnamefont {N.}~\bibnamefont {Dan'shin}}, \bibinfo {author} {\bibfnamefont {Y.~I.}\ \bibnamefont {Nepochatykh}}, \ and\ \bibinfo {author} {\bibfnamefont {V.}~\bibnamefont {Shkar}},\ }\bibfield  {title} {\enquote {\bibinfo {title} {The role of precession and longitudinal oscillations of the magnetization in the spin reorientation dynamics of fe 3 bo 6},}\ }\href@noop {} {\bibfield  {journal} {\bibinfo  {journal} {Soviet Journal of Experimental and Theoretical Physics}\ }\textbf {\bibinfo {volume} {82}},\ \bibinfo {pages} {341--346} (\bibinfo {year} {1996})}\BibitemShut {NoStop}%
\bibitem [{\citenamefont {Buchel’nikov}\ \emph {et~al.}(2001)\citenamefont {Buchel’nikov}, \citenamefont {Dan’shin}, \citenamefont {Nepochatykh},\ and\ \citenamefont {Shavrov}}]{buchel2001observation}%
  \BibitemOpen
  \bibfield  {author} {\bibinfo {author} {\bibfnamefont {V.}~\bibnamefont {Buchel’nikov}}, \bibinfo {author} {\bibfnamefont {N.}~\bibnamefont {Dan’shin}}, \bibinfo {author} {\bibfnamefont {Y.~I.}\ \bibnamefont {Nepochatykh}}, \ and\ \bibinfo {author} {\bibfnamefont {V.}~\bibnamefont {Shavrov}},\ }\bibfield  {title} {\enquote {\bibinfo {title} {The observation of the contribution of longitudinal susceptibility to the frequency of the soft magnetoresonance mode in smfeo 3},}\ }\href@noop {} {\bibfield  {journal} {\bibinfo  {journal} {Journal of Experimental and Theoretical Physics}\ }\textbf {\bibinfo {volume} {92}},\ \bibinfo {pages} {634--642} (\bibinfo {year} {2001})}\BibitemShut {NoStop}%
\bibitem [{\citenamefont {Dan'shin}\ and\ \citenamefont {Nepochatykh}(1998)}]{dan1998contribution}%
  \BibitemOpen
  \bibfield  {author} {\bibinfo {author} {\bibfnamefont {N.}~\bibnamefont {Dan'shin}}\ and\ \bibinfo {author} {\bibfnamefont {Y.~I.}\ \bibnamefont {Nepochatykh}},\ }\bibfield  {title} {\enquote {\bibinfo {title} {The contribution of longitudinal vibrations of magnetization to the spin dynamics of spontaneous reorientation},}\ }\href@noop {} {\bibfield  {journal} {\bibinfo  {journal} {Fizika Nizkikh Temperatur}\ }\textbf {\bibinfo {volume} {24}},\ \bibinfo {pages} {353--359} (\bibinfo {year} {1998})}\BibitemShut {NoStop}%
\bibitem [{\citenamefont {Dan'shin}\ and\ \citenamefont {Kramarchuk}(1993)}]{dan1993energy}%
  \BibitemOpen
  \bibfield  {author} {\bibinfo {author} {\bibfnamefont {N.}~\bibnamefont {Dan'shin}}\ and\ \bibinfo {author} {\bibfnamefont {G.}~\bibnamefont {Kramarchuk}},\ }\bibfield  {title} {\enquote {\bibinfo {title} {Energy gaps in spin-wave spectrum of rare-earth orthoferrites in magnetic field},}\ }\href@noop {} {\bibfield  {journal} {\bibinfo  {journal} {Fizika Nizkikh Temperatur}\ }\textbf {\bibinfo {volume} {19}},\ \bibinfo {pages} {888--894} (\bibinfo {year} {1993})}\BibitemShut {NoStop}%
\bibitem [{\citenamefont {Melcher}(1970)}]{melcher1970magnon}%
  \BibitemOpen
  \bibfield  {author} {\bibinfo {author} {\bibfnamefont {R.}~\bibnamefont {Melcher}},\ }\bibfield  {title} {\enquote {\bibinfo {title} {Magnon-phonon interactions in mnf2},}\ }\href@noop {} {\bibfield  {journal} {\bibinfo  {journal} {Journal of Applied Physics}\ }\textbf {\bibinfo {volume} {41}},\ \bibinfo {pages} {1412--1414} (\bibinfo {year} {1970})}\BibitemShut {NoStop}%
\bibitem [{\citenamefont {Lee}\ and\ \citenamefont {Teale}(1979)}]{lee1979ferromagnetic}%
  \BibitemOpen
  \bibfield  {author} {\bibinfo {author} {\bibfnamefont {E.}~\bibnamefont {Lee}}\ and\ \bibinfo {author} {\bibfnamefont {R.}~\bibnamefont {Teale}},\ }\bibfield  {title} {\enquote {\bibinfo {title} {Ferromagnetic resonance in the frozen lattice model},}\ }\href@noop {} {\bibfield  {journal} {\bibinfo  {journal} {Journal of Physics C: Solid State Physics}\ }\textbf {\bibinfo {volume} {12}},\ \bibinfo {pages} {1131} (\bibinfo {year} {1979})}\BibitemShut {NoStop}%
\bibitem [{\citenamefont {Konstantinov}\ and\ \citenamefont {Perel}(1960)}]{konstantinov1960possibility}%
  \BibitemOpen
  \bibfield  {author} {\bibinfo {author} {\bibfnamefont {O.}~\bibnamefont {Konstantinov}}\ and\ \bibinfo {author} {\bibfnamefont {V.}~\bibnamefont {Perel}},\ }\bibfield  {title} {\enquote {\bibinfo {title} {On the possibility of passage of electromagnetic waves through metals in a strong magnetic field},}\ }\href@noop {} {\bibfield  {journal} {\bibinfo  {journal} {Zhur. Eksptl'. i Teoret. Fiz.}\ }\textbf {\bibinfo {volume} {38}} (\bibinfo {year} {1960})}\BibitemShut {NoStop}%
\bibitem [{\citenamefont {Abrikosov}(2017)}]{abrikosov2017fundamentals}%
  \BibitemOpen
  \bibfield  {author} {\bibinfo {author} {\bibfnamefont {A.~A.}\ \bibnamefont {Abrikosov}},\ }\href@noop {} {\emph {\bibinfo {title} {Fundamentals of the Theory of Metals}}}\ (\bibinfo  {publisher} {Courier Dover Publications},\ \bibinfo {year} {2017})\BibitemShut {NoStop}%
\bibitem [{\citenamefont {Stern}\ and\ \citenamefont {Callen}(1963)}]{stern1963helicons}%
  \BibitemOpen
  \bibfield  {author} {\bibinfo {author} {\bibfnamefont {E.~A.}\ \bibnamefont {Stern}}\ and\ \bibinfo {author} {\bibfnamefont {E.~R.}\ \bibnamefont {Callen}},\ }\bibfield  {title} {\enquote {\bibinfo {title} {Helicons and magnons in magnetically ordered conductors},}\ }\href@noop {} {\bibfield  {journal} {\bibinfo  {journal} {Physical Review}\ }\textbf {\bibinfo {volume} {131}},\ \bibinfo {pages} {512} (\bibinfo {year} {1963})}\BibitemShut {NoStop}%
\bibitem [{\citenamefont {Spector}\ and\ \citenamefont {Casselman}(1965)}]{spector1965interaction}%
  \BibitemOpen
  \bibfield  {author} {\bibinfo {author} {\bibfnamefont {H.}~\bibnamefont {Spector}}\ and\ \bibinfo {author} {\bibfnamefont {T.}~\bibnamefont {Casselman}},\ }\bibfield  {title} {\enquote {\bibinfo {title} {Interaction of alfven waves and spin waves in a ferromagnetic metal},}\ }\href@noop {} {\bibfield  {journal} {\bibinfo  {journal} {Physical Review}\ }\textbf {\bibinfo {volume} {139}},\ \bibinfo {pages} {A1594} (\bibinfo {year} {1965})}\BibitemShut {NoStop}%
\bibitem [{\citenamefont {Bar’yakhtar}, \citenamefont {Savchenko},\ and\ \citenamefont {Stepanov}(1966)}]{bar1966interaction}%
  \BibitemOpen
  \bibfield  {author} {\bibinfo {author} {\bibfnamefont {V.}~\bibnamefont {Bar’yakhtar}}, \bibinfo {author} {\bibfnamefont {M.}~\bibnamefont {Savchenko}}, \ and\ \bibinfo {author} {\bibfnamefont {K.}~\bibnamefont {Stepanov}},\ }\bibfield  {title} {\enquote {\bibinfo {title} {Interaction of plasma and spin waves in ferromagnetic semiconductors and metals},}\ }\href@noop {} {\bibfield  {journal} {\bibinfo  {journal} {Soviet Physics JETP}\ }\textbf {\bibinfo {volume} {23}},\ \bibinfo {pages} {383--390} (\bibinfo {year} {1966})}\BibitemShut {NoStop}%
\bibitem [{\citenamefont {Buchel'nikov}, \citenamefont {Bychkov},\ and\ \citenamefont {Shavrov}(1988)}]{buche88fmm}%
  \BibitemOpen
  \bibfield  {author} {\bibinfo {author} {\bibfnamefont {V.}~\bibnamefont {Buchel'nikov}}, \bibinfo {author} {\bibfnamefont {I.}~\bibnamefont {Bychkov}}, \ and\ \bibinfo {author} {\bibfnamefont {V.}~\bibnamefont {Shavrov}},\ }\bibfield  {title} {\enquote {\bibinfo {title} {Coupled magnetoelastic and electromagnetic waves in magnets near orientational phase transition points},}\ }\href@noop {} {\bibfield  {journal} {\bibinfo  {journal} {The Physics of Metals and Metallography}\ }\textbf {\bibinfo {volume} {66}},\ \bibinfo {pages} {9--13} (\bibinfo {year} {1988})}\BibitemShut {NoStop}%
\bibitem [{\citenamefont {Skobov}\ and\ \citenamefont {Kaner}(1964)}]{skobov1964theory}%
  \BibitemOpen
  \bibfield  {author} {\bibinfo {author} {\bibfnamefont {V.}~\bibnamefont {Skobov}}\ and\ \bibinfo {author} {\bibfnamefont {E.}~\bibnamefont {Kaner}},\ }\bibfield  {title} {\enquote {\bibinfo {title} {Theory of coupled electromagnetic and acoustic waves in metals in a magnetic field},}\ }\href@noop {} {\bibfield  {journal} {\bibinfo  {journal} {Soviet Physics JETP}\ }\textbf {\bibinfo {volume} {19}} (\bibinfo {year} {1964})}\BibitemShut {NoStop}%
\bibitem [{\citenamefont {Blank}\ and\ \citenamefont {Kaganov}(1968)}]{blank1968ferromagnetic}%
  \BibitemOpen
  \bibfield  {author} {\bibinfo {author} {\bibfnamefont {A.~Y.}\ \bibnamefont {Blank}}\ and\ \bibinfo {author} {\bibfnamefont {M.~I.}\ \bibnamefont {Kaganov}},\ }\bibfield  {title} {\enquote {\bibinfo {title} {Ferromagnetic resonance and plasma effects in metals},}\ }\href@noop {} {\bibfield  {journal} {\bibinfo  {journal} {Soviet Physics Uspekhi}\ }\textbf {\bibinfo {volume} {10}},\ \bibinfo {pages} {536} (\bibinfo {year} {1968})}\BibitemShut {NoStop}%
\bibitem [{\citenamefont {Turov}\ and\ \citenamefont {Kaibichev}(1989)}]{turov1989ground}%
  \BibitemOpen
  \bibfield  {author} {\bibinfo {author} {\bibfnamefont {E.~A.}\ \bibnamefont {Turov}}\ and\ \bibinfo {author} {\bibfnamefont {I.}~\bibnamefont {Kaibichev}},\ }\bibfield  {title} {\enquote {\bibinfo {title} {Ground state stability and faraday acoustic effect in a ferromagnet. rotationally invariant theory},}\ }\href@noop {} {\bibfield  {journal} {\bibinfo  {journal} {Fizika Tverdogo Tela}\ }\textbf {\bibinfo {volume} {31}},\ \bibinfo {pages} {138--143} (\bibinfo {year} {1989})}\BibitemShut {NoStop}%
\bibitem [{\citenamefont {Kaibichev}\ and\ \citenamefont {Shavrov}(1993)}]{kaibichev1993rotation}%
  \BibitemOpen
  \bibfield  {author} {\bibinfo {author} {\bibfnamefont {I.}~\bibnamefont {Kaibichev}}\ and\ \bibinfo {author} {\bibfnamefont {V.}~\bibnamefont {Shavrov}},\ }\bibfield  {title} {\enquote {\bibinfo {title} {Rotation-invariant theory of acoustic birefringence in a ferromagnet},}\ }\href@noop {} {\bibfield  {journal} {\bibinfo  {journal} {Acoustical physics}\ }\textbf {\bibinfo {volume} {39}},\ \bibinfo {pages} {353--355} (\bibinfo {year} {1993})}\BibitemShut {NoStop}%
\bibitem [{\citenamefont {Kaibichev}(1993)}]{kaibichev1993stability}%
  \BibitemOpen
  \bibfield  {author} {\bibinfo {author} {\bibfnamefont {I.}~\bibnamefont {Kaibichev}},\ }\bibfield  {title} {\enquote {\bibinfo {title} {Stability of the ground state, spectra of transverse magnetoelastic waves, and the acoustic faraday effect in tetragonal antiferromagnets. rotationally invariant theory},}\ }\href@noop {} {\bibfield  {journal} {\bibinfo  {journal} {Physics of the solid state}\ }\textbf {\bibinfo {volume} {35}},\ \bibinfo {pages} {76--81} (\bibinfo {year} {1993})}\BibitemShut {NoStop}%
\bibitem [{\citenamefont {Buchel'nikov}\ \emph {et~al.}(2000)\citenamefont {Buchel'nikov}, \citenamefont {Babushkin}, \citenamefont {Bychkov},\ and\ \citenamefont {Shavrov}}]{buche2000fmm}%
  \BibitemOpen
  \bibfield  {author} {\bibinfo {author} {\bibfnamefont {V.}~\bibnamefont {Buchel'nikov}}, \bibinfo {author} {\bibfnamefont {A.}~\bibnamefont {Babushkin}}, \bibinfo {author} {\bibfnamefont {I.}~\bibnamefont {Bychkov}}, \ and\ \bibinfo {author} {\bibfnamefont {V.}~\bibnamefont {Shavrov}},\ }\bibfield  {title} {\enquote {\bibinfo {title} {Coupled magnetoelastic and electromagnetic waves in cubic ferromagnets in the region of reorientational phase transitions},}\ }\href@noop {} {\bibfield  {journal} {\bibinfo  {journal} {The Physics of Metals and Metallography}\ }\textbf {\bibinfo {volume} {90}},\ \bibinfo {pages} {323--329} (\bibinfo {year} {2000})}\BibitemShut {NoStop}%
\bibitem [{\citenamefont {Novotny}(2010)}]{novotny2010}%
  \BibitemOpen
  \bibfield  {author} {\bibinfo {author} {\bibfnamefont {L.}~\bibnamefont {Novotny}},\ }\bibfield  {title} {\enquote {\bibinfo {title} {Strong coupling, energy splitting, and level crossings: A classical perspective},}\ }\href {\doibase 10.1119/1.3471177} {\bibfield  {journal} {\bibinfo  {journal} {American Journal of Physics}\ }\textbf {\bibinfo {volume} {78}},\ \bibinfo {pages} {1199--1202} (\bibinfo {year} {2010})},\ \Eprint {http://arxiv.org/abs/https://pubs.aip.org/aapt/ajp/article-pdf/78/11/1199/13132946/1199\_1\_online.pdf} {https://pubs.aip.org/aapt/ajp/article-pdf/78/11/1199/13132946/1199\_1\_online.pdf} \BibitemShut {NoStop}%
\bibitem [{\citenamefont {Rodriguez}(2016)}]{rodriguez2016}%
  \BibitemOpen
  \bibfield  {author} {\bibinfo {author} {\bibfnamefont {S.~R.-K.}\ \bibnamefont {Rodriguez}},\ }\bibfield  {title} {\enquote {\bibinfo {title} {Classical and quantum distinctions between weak and strong coupling},}\ }\href {\doibase 10.1088/0143-0807/37/2/025802} {\bibfield  {journal} {\bibinfo  {journal} {European Journal of Physics}\ }\textbf {\bibinfo {volume} {37}},\ \bibinfo {pages} {025802} (\bibinfo {year} {2016})}\BibitemShut {NoStop}%
\bibitem [{\citenamefont {Holanda}\ \emph {et~al.}(2018)\citenamefont {Holanda}, \citenamefont {Maior}, \citenamefont {Azevedo},\ and\ \citenamefont {Rezende}}]{holanda2018}%
  \BibitemOpen
  \bibfield  {author} {\bibinfo {author} {\bibfnamefont {J.}~\bibnamefont {Holanda}}, \bibinfo {author} {\bibfnamefont {D.~S.}\ \bibnamefont {Maior}}, \bibinfo {author} {\bibfnamefont {A.}~\bibnamefont {Azevedo}}, \ and\ \bibinfo {author} {\bibfnamefont {S.~M.}\ \bibnamefont {Rezende}},\ }\bibfield  {title} {\enquote {\bibinfo {title} {Detecting the phonon spin in magnon-phonon conversion experiments},}\ }\href {\doibase 10.1038/s41567-018-0079-y} {\bibfield  {journal} {\bibinfo  {journal} {Nature Physics}\ }\textbf {\bibinfo {volume} {14}},\ \bibinfo {pages} {500--506} (\bibinfo {year} {2018})}\BibitemShut {NoStop}%
\bibitem [{\citenamefont {Man}\ \emph {et~al.}(2017)\citenamefont {Man}, \citenamefont {Shi}, \citenamefont {Xu}, \citenamefont {Xu}, \citenamefont {Chen}, \citenamefont {Sullivan}, \citenamefont {Zhou}, \citenamefont {Xia}, \citenamefont {Shi},\ and\ \citenamefont {Dai}}]{man2017}%
  \BibitemOpen
  \bibfield  {author} {\bibinfo {author} {\bibfnamefont {H.}~\bibnamefont {Man}}, \bibinfo {author} {\bibfnamefont {Z.}~\bibnamefont {Shi}}, \bibinfo {author} {\bibfnamefont {G.}~\bibnamefont {Xu}}, \bibinfo {author} {\bibfnamefont {Y.}~\bibnamefont {Xu}}, \bibinfo {author} {\bibfnamefont {X.}~\bibnamefont {Chen}}, \bibinfo {author} {\bibfnamefont {S.}~\bibnamefont {Sullivan}}, \bibinfo {author} {\bibfnamefont {J.}~\bibnamefont {Zhou}}, \bibinfo {author} {\bibfnamefont {K.}~\bibnamefont {Xia}}, \bibinfo {author} {\bibfnamefont {J.}~\bibnamefont {Shi}}, \ and\ \bibinfo {author} {\bibfnamefont {P.}~\bibnamefont {Dai}},\ }\bibfield  {title} {\enquote {\bibinfo {title} {Direct observation of magnon-phonon coupling in yttrium iron garnet},}\ }\href {\doibase 10.1103/PhysRevB.96.100406} {\bibfield  {journal} {\bibinfo  {journal} {Phys. Rev. B}\ }\textbf {\bibinfo {volume} {96}},\ \bibinfo {pages} {100406} (\bibinfo {year} {2017})}\BibitemShut {NoStop}%
\bibitem [{\citenamefont {Vogel}\ \emph {et~al.}(2015)\citenamefont {Vogel}, \citenamefont {Chumak}, \citenamefont {Waller}, \citenamefont {Langner}, \citenamefont {Vasyuchka}, \citenamefont {Hillebrands},\ and\ \citenamefont {von Freymann}}]{vogel2015}%
  \BibitemOpen
  \bibfield  {author} {\bibinfo {author} {\bibfnamefont {M.}~\bibnamefont {Vogel}}, \bibinfo {author} {\bibfnamefont {A.~V.}\ \bibnamefont {Chumak}}, \bibinfo {author} {\bibfnamefont {E.~H.}\ \bibnamefont {Waller}}, \bibinfo {author} {\bibfnamefont {T.}~\bibnamefont {Langner}}, \bibinfo {author} {\bibfnamefont {V.~I.}\ \bibnamefont {Vasyuchka}}, \bibinfo {author} {\bibfnamefont {B.}~\bibnamefont {Hillebrands}}, \ and\ \bibinfo {author} {\bibfnamefont {G.}~\bibnamefont {von Freymann}},\ }\bibfield  {title} {\enquote {\bibinfo {title} {Optically reconfigurable magnetic materials},}\ }\href {\doibase 10.1038/nphys3325} {\bibfield  {journal} {\bibinfo  {journal} {Nature Physics}\ }\textbf {\bibinfo {volume} {11}},\ \bibinfo {pages} {487--491} (\bibinfo {year} {2015})}\BibitemShut {NoStop}%
\bibitem [{\citenamefont {Krawczyk}\ and\ \citenamefont {Grundler}(2014)}]{krawczyk2014}%
  \BibitemOpen
  \bibfield  {author} {\bibinfo {author} {\bibfnamefont {M.}~\bibnamefont {Krawczyk}}\ and\ \bibinfo {author} {\bibfnamefont {D.}~\bibnamefont {Grundler}},\ }\bibfield  {title} {\enquote {\bibinfo {title} {Review and prospects of magnonic crystals and devices with reprogrammable band structure},}\ }\href {\doibase 10.1088/0953-8984/26/12/123202} {\bibfield  {journal} {\bibinfo  {journal} {Journal of Physics: Condensed Matter}\ }\textbf {\bibinfo {volume} {26}},\ \bibinfo {pages} {123202} (\bibinfo {year} {2014})}\BibitemShut {NoStop}%
\bibitem [{\citenamefont {Godejohann}\ \emph {et~al.}(2020)\citenamefont {Godejohann}, \citenamefont {Scherbakov}, \citenamefont {Kukhtaruk}, \citenamefont {Poddubny}, \citenamefont {Yaremkevich}, \citenamefont {Wang}, \citenamefont {Nadzeyka}, \citenamefont {Yakovlev}, \citenamefont {Rushforth}, \citenamefont {Akimov},\ and\ \citenamefont {Bayer}}]{godejohann2020}%
  \BibitemOpen
  \bibfield  {author} {\bibinfo {author} {\bibfnamefont {F.}~\bibnamefont {Godejohann}}, \bibinfo {author} {\bibfnamefont {A.~V.}\ \bibnamefont {Scherbakov}}, \bibinfo {author} {\bibfnamefont {S.~M.}\ \bibnamefont {Kukhtaruk}}, \bibinfo {author} {\bibfnamefont {A.~N.}\ \bibnamefont {Poddubny}}, \bibinfo {author} {\bibfnamefont {D.~D.}\ \bibnamefont {Yaremkevich}}, \bibinfo {author} {\bibfnamefont {M.}~\bibnamefont {Wang}}, \bibinfo {author} {\bibfnamefont {A.}~\bibnamefont {Nadzeyka}}, \bibinfo {author} {\bibfnamefont {D.~R.}\ \bibnamefont {Yakovlev}}, \bibinfo {author} {\bibfnamefont {A.~W.}\ \bibnamefont {Rushforth}}, \bibinfo {author} {\bibfnamefont {A.~V.}\ \bibnamefont {Akimov}}, \ and\ \bibinfo {author} {\bibfnamefont {M.}~\bibnamefont {Bayer}},\ }\bibfield  {title} {\enquote {\bibinfo {title} {Magnon polaron formed by selectively coupled coherent magnon and phonon modes of a surface patterned ferromagnet},}\ }\href {\doibase 10.1103/PhysRevB.102.144438} {\bibfield  {journal} {\bibinfo  {journal} {Phys.
  Rev. B}\ }\textbf {\bibinfo {volume} {102}},\ \bibinfo {pages} {144438} (\bibinfo {year} {2020})}\BibitemShut {NoStop}%
\bibitem [{\citenamefont {Berk}\ \emph {et~al.}(2019)\citenamefont {Berk}, \citenamefont {Jaris}, \citenamefont {Yang}, \citenamefont {Dhuey}, \citenamefont {Cabrini},\ and\ \citenamefont {Schmidt}}]{berk2019}%
  \BibitemOpen
  \bibfield  {author} {\bibinfo {author} {\bibfnamefont {C.}~\bibnamefont {Berk}}, \bibinfo {author} {\bibfnamefont {M.}~\bibnamefont {Jaris}}, \bibinfo {author} {\bibfnamefont {W.}~\bibnamefont {Yang}}, \bibinfo {author} {\bibfnamefont {S.}~\bibnamefont {Dhuey}}, \bibinfo {author} {\bibfnamefont {S.}~\bibnamefont {Cabrini}}, \ and\ \bibinfo {author} {\bibfnamefont {H.}~\bibnamefont {Schmidt}},\ }\bibfield  {title} {\enquote {\bibinfo {title} {Strongly coupled magnon-phonon dynamics in a single nanomagnet},}\ }\href {\doibase 10.1038/s41467-019-10545-x} {\bibfield  {journal} {\bibinfo  {journal} {Nature Communications}\ }\textbf {\bibinfo {volume} {10}},\ \bibinfo {pages} {2652} (\bibinfo {year} {2019})}\BibitemShut {NoStop}%
\bibitem [{\citenamefont {Dan'shin}, \citenamefont {Kovtun},\ and\ \citenamefont {Sdvizhkov}(1986)}]{danshin86mhd}%
  \BibitemOpen
  \bibfield  {author} {\bibinfo {author} {\bibfnamefont {N.}~\bibnamefont {Dan'shin}}, \bibinfo {author} {\bibfnamefont {N.}~\bibnamefont {Kovtun}}, \ and\ \bibinfo {author} {\bibfnamefont {M.}~\bibnamefont {Sdvizhkov}},\ }\bibfield  {title} {\enquote {\bibinfo {title} {Magnetohydrodynamical resonance in the region of low-temperature phase transition in $\mathrm{ErFeO}_{3}$},}\ }\href@noop {} {\bibfield  {journal} {\bibinfo  {journal} {Fizika Tverdogo Tela}\ }\textbf {\bibinfo {volume} {28}},\ \bibinfo {pages} {1200--1202} (\bibinfo {year} {1986})}\BibitemShut {NoStop}%
\bibitem [{\citenamefont {Jiles}(2015)}]{jiles2015introduction}%
  \BibitemOpen
  \bibfield  {author} {\bibinfo {author} {\bibfnamefont {D.}~\bibnamefont {Jiles}},\ }\href@noop {} {\emph {\bibinfo {title} {Introduction to magnetism and magnetic materials}}}\ (\bibinfo  {publisher} {CRC press},\ \bibinfo {year} {2015})\BibitemShut {NoStop}%
\bibitem [{\citenamefont {Smolenski{\u\i}}\ and\ \citenamefont {Chupis}(1982)}]{smolenskiui1982ferroelectromagnets}%
  \BibitemOpen
  \bibfield  {author} {\bibinfo {author} {\bibfnamefont {G.}~\bibnamefont {Smolenski{\u\i}}}\ and\ \bibinfo {author} {\bibfnamefont {I.~E.}\ \bibnamefont {Chupis}},\ }\bibfield  {title} {\enquote {\bibinfo {title} {Ferroelectromagnets},}\ }\href@noop {} {\bibfield  {journal} {\bibinfo  {journal} {Soviet Physics Uspekhi}\ }\textbf {\bibinfo {volume} {25}},\ \bibinfo {pages} {475} (\bibinfo {year} {1982})}\BibitemShut {NoStop}%
\bibitem [{\citenamefont {Schmid}(1994)}]{schmid1994multi}%
  \BibitemOpen
  \bibfield  {author} {\bibinfo {author} {\bibfnamefont {H.}~\bibnamefont {Schmid}},\ }\bibfield  {title} {\enquote {\bibinfo {title} {Multi-ferroic magnetoelectrics},}\ }\href@noop {} {\bibfield  {journal} {\bibinfo  {journal} {Ferroelectrics}\ }\textbf {\bibinfo {volume} {162}},\ \bibinfo {pages} {317--338} (\bibinfo {year} {1994})}\BibitemShut {NoStop}%
\bibitem [{\citenamefont {Izyumov}\ and\ \citenamefont {Skryabin}(2001)}]{izyumov2001double}%
  \BibitemOpen
  \bibfield  {author} {\bibinfo {author} {\bibfnamefont {Y.~A.}\ \bibnamefont {Izyumov}}\ and\ \bibinfo {author} {\bibfnamefont {Y.~N.}\ \bibnamefont {Skryabin}},\ }\bibfield  {title} {\enquote {\bibinfo {title} {Double exchange model and the unique properties of the manganites},}\ }\href@noop {} {\bibfield  {journal} {\bibinfo  {journal} {Physics-Uspekhi}\ }\textbf {\bibinfo {volume} {44}},\ \bibinfo {pages} {109} (\bibinfo {year} {2001})}\BibitemShut {NoStop}%
\bibitem [{\citenamefont {Mukherjee}\ \emph {et~al.}(2013)\citenamefont {Mukherjee}, \citenamefont {Cole}, \citenamefont {Woodward}, \citenamefont {Randeria},\ and\ \citenamefont {Trivedi}}]{mukherjee2013theory}%
  \BibitemOpen
  \bibfield  {author} {\bibinfo {author} {\bibfnamefont {A.}~\bibnamefont {Mukherjee}}, \bibinfo {author} {\bibfnamefont {W.~S.}\ \bibnamefont {Cole}}, \bibinfo {author} {\bibfnamefont {P.}~\bibnamefont {Woodward}}, \bibinfo {author} {\bibfnamefont {M.}~\bibnamefont {Randeria}}, \ and\ \bibinfo {author} {\bibfnamefont {N.}~\bibnamefont {Trivedi}},\ }\bibfield  {title} {\enquote {\bibinfo {title} {Theory of strain-controlled magnetotransport and stabilization of the ferromagnetic insulating phase in manganite thin films},}\ }\href@noop {} {\bibfield  {journal} {\bibinfo  {journal} {Physical review letters}\ }\textbf {\bibinfo {volume} {110}},\ \bibinfo {pages} {157201} (\bibinfo {year} {2013})}\BibitemShut {NoStop}%
\bibitem [{\citenamefont {Hatano}\ \emph {et~al.}(2014)\citenamefont {Hatano}, \citenamefont {Sheng}, \citenamefont {Nakamura}, \citenamefont {Nakano}, \citenamefont {Kawasaki}, \citenamefont {Iwasa},\ and\ \citenamefont {Tokura}}]{hatano2014gate}%
  \BibitemOpen
  \bibfield  {author} {\bibinfo {author} {\bibfnamefont {T.}~\bibnamefont {Hatano}}, \bibinfo {author} {\bibfnamefont {Z.}~\bibnamefont {Sheng}}, \bibinfo {author} {\bibfnamefont {M.}~\bibnamefont {Nakamura}}, \bibinfo {author} {\bibfnamefont {M.}~\bibnamefont {Nakano}}, \bibinfo {author} {\bibfnamefont {M.}~\bibnamefont {Kawasaki}}, \bibinfo {author} {\bibfnamefont {Y.}~\bibnamefont {Iwasa}}, \ and\ \bibinfo {author} {\bibfnamefont {Y.}~\bibnamefont {Tokura}},\ }\bibfield  {title} {\enquote {\bibinfo {title} {Gate control of percolative conduction in strongly correlated manganite films.}}\ }\href@noop {} {\bibfield  {journal} {\bibinfo  {journal} {Advanced Materials (Deerfield Beach, Fla.)}\ }\textbf {\bibinfo {volume} {26}},\ \bibinfo {pages} {2874--2877} (\bibinfo {year} {2014})}\BibitemShut {NoStop}%
\bibitem [{\citenamefont {Bar'Yakhtar}, \citenamefont {Savchenko},\ and\ \citenamefont {Tarasenko}(1966)}]{bar1966coupled}%
  \BibitemOpen
  \bibfield  {author} {\bibinfo {author} {\bibfnamefont {V.}~\bibnamefont {Bar'Yakhtar}}, \bibinfo {author} {\bibfnamefont {M.}~\bibnamefont {Savchenko}}, \ and\ \bibinfo {author} {\bibfnamefont {V.}~\bibnamefont {Tarasenko}},\ }\bibfield  {title} {\enquote {\bibinfo {title} {Coupled magnetoelastic waves in antiferromagnets in strong magnetic fields},}\ }\href@noop {} {\bibfield  {journal} {\bibinfo  {journal} {Soviet Physics JETP}\ }\textbf {\bibinfo {volume} {22}} (\bibinfo {year} {1966})}\BibitemShut {NoStop}%
\bibitem [{\citenamefont {Liebermann}\ and\ \citenamefont {Banerjee}(1970)}]{liebermann1970anomalies}%
  \BibitemOpen
  \bibfield  {author} {\bibinfo {author} {\bibfnamefont {R.~C.}\ \bibnamefont {Liebermann}}\ and\ \bibinfo {author} {\bibfnamefont {S.~K.}\ \bibnamefont {Banerjee}},\ }\bibfield  {title} {\enquote {\bibinfo {title} {Anomalies in the compressional and shear properties of hematite in the region of the morin transition},}\ }\href@noop {} {\bibfield  {journal} {\bibinfo  {journal} {Journal of Applied Physics}\ }\textbf {\bibinfo {volume} {41}},\ \bibinfo {pages} {1414--1416} (\bibinfo {year} {1970})}\BibitemShut {NoStop}%
\end{thebibliography}%
\end{document}